\documentclass[fleqn,usenatbib]{mnras}
\usepackage{hyperref}

\usepackage[dvipsnames]{xcolor}
\usepackage{newtxtext,newtxmath,bm,mathtools,placeins,tikz}
\usetikzlibrary{shapes}

\usepackage[T1]{fontenc}
\usepackage[utf8]{inputenc}
\usepackage{ae,aecompl}

\usepackage{todonotes}

\usepackage{graphicx}	
\usepackage{amsmath,siunitx}	
\usepackage{amssymb}	

\newcommand{\mmat}[1]{{\mathbf{#1}}}
\newcommand{\mvec}[1]{{\bm{#1}}}

\newcommand{\sdssiii}{{SDSS-III/BOSS}}
\newcommand{\borg}{\textsc{borg}}
\newcommand{\healpix}{\textsc{HEALPix}}
\newcommand{\Mpch}{$h^{-1}$~Mpc}
\newcommand{\hMpc}{$h$~Mpc$^{-1}$}
\newcommand{\lcdm}{$\Lambda$CDM}
\newcommand{\kmsMpc}{km~s$^{-1}$~Mpc$^{-1}$}

\newcommand{\proba}{P}

\DeclarePairedDelimiter\floor{\lfloor}{\rfloor}

\definecolor{darkgreen}{rgb}{0.0,0.5,0.0}

\title[Cosmological density inference with SDSS3-BOSS]{Systematic-free inference of the cosmic matter density field from SDSS3-BOSS data}

\author[G. Lavaux, J. Jasche \& F. Leclercq]{
Guilhem Lavaux,$^{1}$\thanks{E-mail: \href{mailto:guilhem.lavaux@iap.fr}{guilhem.lavaux@iap.fr}}
Jens Jasche$^{2}$ and
Florent Leclercq$^{3}$
\\
$^{1}$ CNRS \& Sorbonne Universit\'{e}, UMR7095, Institut d'Astrophysique de Paris, F-75014, Paris, France\\
$^{2}$ The Oskar Klein Centre, Department of Physics, Stockholm University, AlbaNova University Centre,
SE 106 91 Stockholm, Sweden \\
$^{3}$ Imperial Centre for Inference and Cosmology (ICIC) \& Astrophysics Group, Imperial College London,\\ Blackett Laboratory, Prince Consort Road, London SW7 2AZ, United Kingdom
}

\date{Accepted XXX. Received YYY; in original form ZZZ}

\pubyear{2019}

\begin{document}
\label{firstpage}
\pagerange{\pageref{firstpage}--\pageref{lastpage}}
\todototoc

\maketitle

\begin{abstract}
We perform an analysis of the three-dimensional cosmic matter density field traced by galaxies of the \sdssiii{} galaxy sample. The systematic-free nature of this analysis is confirmed by two elements: the successful cross-correlation with the gravitational lensing observations derived from \emph{Planck 2018} data and the absence of bias at scales $k\simeq 10^{-3}-10^{-2}${\hMpc} in the \textit{a posteriori} power spectrum of recovered initial conditions. 
Our analysis builds upon our algorithm for Bayesian Origin Reconstruction from Galaxies (\borg{}) and uses a physical model of cosmic structure formation to infer physically meaningful cosmic structures and their corresponding dynamics from deep galaxy observations. Our approach accounts for redshift-space distortions and light-cone effects inherent to deep observations. We also apply detailed corrections to account for known and unknown foreground contaminations, selection effects and galaxy biases. 
We obtain maps of residual, so far unexplained, systematic effects in the spectroscopic data of \sdssiii{}. Our results show that unbiased and physically plausible models of the cosmic large scale structure can be obtained from present and next-generation galaxy surveys.
\end{abstract}




\section{Introduction}

The measurement of clustering properties with modern galaxy surveys is achieving unprecedented precision on many scales of interest for cosmology \citep{Ross2017_BOSS}. However, even for well-controlled galaxy surveys such as \sdssiii{} \citep{Einstein2011_SDSS3}, systematic effects affect the largest spatial scales. This is clearly illustrated by the recent work of \citet{kalus_map-based_2019}: for scales\footnote{We use $h = H / (100$~\kmsMpc) with $H$ the Hubble constant at redshift $z=0$.} $k\lesssim 10^{-2}${\hMpc} systematics remain a challenge in clustering analyses. Since sampling noise will be reduced with future surveys such as Euclid \citep{Laureijs2011_EUCLID}, the problem will further increase, hampering our capability to do cosmological inference.


Fortunately, over the last decade, Bayesian forward modelling of large-scale structures has come of age and may provide a way out. This method allows, assuming that the initial conditions are drawn statistically fairly from a Gaussian distribution, to model the detail of the observed galaxy distribution. Notably, the {\borg} algorithm \citep{jasche_bayesian_2013} has been successfully applied to the 2M++ galaxy compilation \citep{LH11,lavaux_unmasking_2016,Jasche2019_PM} and to the SDSS-II main galaxy sample
\citep{SDSS7,Jasche2010}. Other groups \citep[in particular the ELUCID projet,][]{Wang2014,Wang2016,Tweed2017} have developed techniques similar in spirit, meeting some success in applying to the SDSS-II main galaxy sample.

In this work, we apply the newly updated \borg{} analysis framework jointly to the two galaxy samples of \sdssiii{}, LOWZ and CMASS. Most analyses run separate analysis on each component, which increases sample variance in their measurement. We are not limited by this aspect and we can add as many surveys as needed, provided that no double counting of a single galaxy occurs. We aim at recovering an unbiased ensemble of history of formation of the large-scale structure, and validate the model with Planck lensing maps \citep{PlanckCollaboration2018a}. Solving this problem will open up new venues to extract cosmological information, in particular with the likelihood-based ALTAIR extension of \borg{} \citep{Ramanah2019} or with the likelihood-free SELFI algorithm \citep{Leclercq2019}.

An important feature of \sdssiii{} is that it provides an excellent test case for the next generation of galaxy surveys, i.e. the Euclid mission \citep{Laureijs2011_EUCLID} and the Large Synoptic Survey Telescope \citep[LSST][]{lsst_science_collaboration_lsst_2009}. It will be impossible to control everything in the data acquisition of these surveys, and the huge number of expected observed galaxies with photometric redshifts ($\sim$20 billions for LSST, $\sim$1 billion for Euclid) will make all classical methods signal-dominated and dangerously sensitive to systematic errors \citep{Laureijs2011_EUCLID,Colavincenzo2017,Monaco2018} to reach sub-percent precision the measurement of the density power spectrum. This further highlights the need to control systematic signals. There is also the interesting possibility that unexploited cosmological signal is available in the data of \sdssiii{}.

To achieve a systematic-free inference we make use of a new likelihood \citep[named ``robust Poisson likelihood'',][]{porqueres_explicit_2019}, and a template matching method for systematics \citep{Jasche2017} to remove the impact of systematic effects on our inference. Usual galaxy survey data analysis rely fully on the existence of maps to correct for large scale, subtle, systematic effects. For example, \citet{leistedt_exploiting_2014} compiled a set of 220 foreground maps of possible contaminants for the inference of the clustering signal of quasars in \sdssiii{}. Later, \citet{Elsner2016,Elsner2017} showed that `extended mode' projection, i.e. the foreground template fitting technique, is almost surely biased, whereas `basic mode' projection is unbiased in most cases. These two approaches resemble what has been done for analysis of Cosmic Microwave Background data obtained from space \citep{Tegmark1997a} and ground observatories \citep[e.g. for ACT and SPT,][]{Fowler2010,Schaffer2011}. In this work, we use the two kind of procedures at the same time.

Another big issue in analysing galaxy surveys is the derivation of the relation between galaxy population and the large scale dark matter field. This relation is generally called the galaxy bias model \citep{Kaiser1984,Desjacques2018}. Typical analysis methods are calibrated on mock data from $N$-body simulations before being applied to galaxy surveys \citep{Chuang2015_EZMOCKS,Kitaura2016_PATCHY,Beutler2017_BOSS,Satpathy2017_BOSS}. A more agnostic procedure would fit this ``bias'' model jointly with the inference of cosmological parameters, the underlying density field and the eventual residual due to systematic effects \citep{Jasche2017}. However, finding a family of bias models sufficiently generic to capture all the unknown small-scale physics, extensible and fast to evaluate, is non-trivial \citep{Schmidt2019}. In this work, we present a novel bias model that has some of these properties.

This paper is organised as follows. In Section~\ref{sec:method}, we present the
overall organisation of the \borg{} inference method, its assumptions and the essential new components of the adopted model to represent the galaxy distribution of \sdssiii{}. In particular, we review the properties of our robust likelihood and introduce a new bias model. Next, we present the pre-processed data provided to the \borg{} inference machine in Section~\ref{sec:data}. We then describe our results on the systematic-free inference of the large-scale structure in Section~\ref{sec:results}, including the systematic maps that we have derived for the \sdssiii{} sample. We conclude in Section~\ref{sec:conclusion}.

\section{Method}
\label{sec:method}

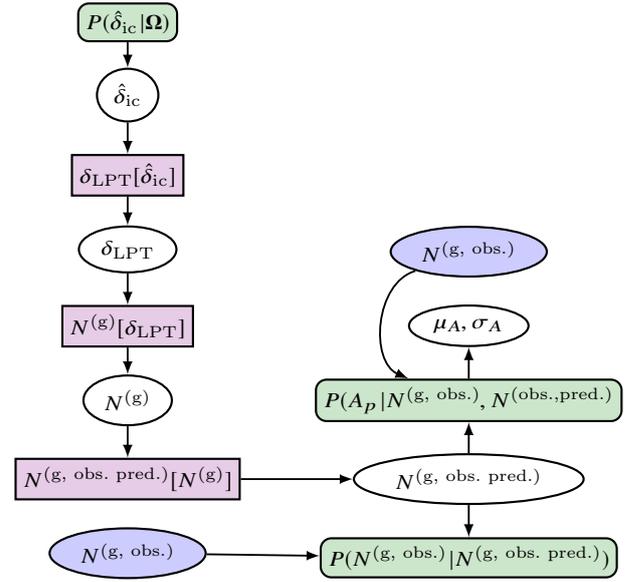
\begin{figure}
  \begin{center}

    \begin{tikzpicture}
    	\pgfdeclarelayer{background}
    	\pgfdeclarelayer{foreground}
    	\pgfsetlayers{background,main,foreground}
    
        \tikzstyle{probability}=[draw, thick, text centered, rounded corners, minimum height=1em, minimum width=1em, fill=darkgreen!20]
    	\tikzstyle{deterministic}=[draw, thick, text centered, minimum height=1.8em, minimum width=1.8em, fill=violet!20]
    	\tikzstyle{variabl}=[ellipse, draw, thick, text centered, minimum height=1em, minimum width=1em]
    	\tikzstyle{data}=[ellipse, draw, thick, text centered, minimum height=1em, minimum width=1em, fill=blue!20]
    
    	\def\blockdist{0.7}
    
        \node (ic) [probability]
        {$\proba(\hat{\delta}_\text{ic}|\mvec{\Omega})$};
        \path (ic.south)+(0,-\blockdist) node (deltaic) [variabl]
        {$\hat{\delta}_\text{ic}$};
        \path (deltaic.south)+(0,-\blockdist) node (deltalpt) [deterministic]
        {$\delta_\text{LPT}[\hat{\delta}_\text{ic}]$};
        \path (deltalpt.south)+(0,-\blockdist) node (deltalpt_field) [variabl]
        {$\delta_\text{LPT}$};
        \path (deltalpt_field.south)+(0,-\blockdist) node (bias) [deterministic]
        {$N^\text{(g)}[\delta_\text{LPT}]$};
        \path (bias.south)+(0,-\blockdist) node (ng_field) [variabl]
        {$N^\text{(g)}$};
        \path (ng_field.south)+(0,-\blockdist) node (obspred) [deterministic]
        {$N^\text{(g, obs. pred.)}[N^\text{(g)}]$};
        \path (obspred.east)+(3,0) node (obspred_field) [variabl]
        {$N^\text{(g, obs. pred.)}$};
        \path (obspred_field.south)+(0,-\blockdist) node (robust) [probability]
        {$\proba(N^\text{(g, obs.)} | N^\text{(g, obs. pred.)})$};
        \path (obspred.south)+(0,-\blockdist) node (obs) [data]
        {$N^\text{(g, obs.)}$};
        \path (obspred_field.north)+(0,\blockdist) node (ap_prob) [probability]
        {$\proba(A_p|N^\text{(g, obs.)},N^\text{(obs.,pred.)}$};
        \path (ap_prob.north)+(0,\blockdist) node (ap_field) [variabl]
        {$\mu_A, \sigma_A$};
        \path (ap_field.north)+(0,\blockdist) node (obs_bis) [data]
        {$N^\text{(g, obs.)}$};
    
    	\path [draw, line width=0.7pt, arrows={-latex}] (ic) -- (deltaic);
    	\path [draw, line width=0.7pt, arrows={-latex}] (deltaic) -- (deltalpt);
    	\path [draw, line width=0.7pt, arrows={-latex}] (deltalpt) -- (deltalpt_field);
    	\path [draw, line width=0.7pt, arrows={-latex}] (deltalpt_field) -- (bias);
    	\path [draw, line width=0.7pt, arrows={-latex}] (bias) -- (ng_field);
    	\path [draw, line width=0.7pt, arrows={-latex}] (ng_field) -- (obspred);
    	\path [draw, line width=0.7pt, arrows={-latex}] (obspred) -- (obspred_field);
    	\path [draw, line width=0.7pt, arrows={-latex}] (obspred_field) -- (robust);
    	\path [draw, line width=0.7pt, arrows={-latex}] (obs) -- (robust);
    	\path [draw, line width=0.7pt, arrows={-latex}] (obspred_field) -- (ap_prob);
    	\path [draw, line width=0.7pt, arrows={-latex}] (ap_prob) -- (ap_field);
    	\draw [line width=0.7pt, -latex] (obs_bis) to [bend right=70] (ap_prob);

    \end{tikzpicture}
  \end{center}
\caption{Hierarchical representation of the Bayesian inference framework used for the analysis of \sdssiii{}. $\mvec{\Omega}$ represents cosmological parameters,  $\hat{\delta}_\text{ic}$ is the set of Fourier modes encoding the initial conditions at $z\simeq 1~000$, $\delta_\text{LPT}$ is the density field obtained from first-order Lagrangian Perturbation Theory, $N^\text{(g)}$ the galaxy number density field derived from the bias model, $N^\text{(g, obs. pred)}$ the galaxy number density field that should be observed after selection and systematic effects, $N^\text{(g, obs.)}$ the actual \sdssiii{} data, $\mu_A$ and $\sigma_A$ the per-pixel mean and standard deviation of the inferred systematic maps. The details are provided in Section~\ref{sec:method}. Purple boxes correspond to a deterministic transition from one field to another. Green boxes are probability distributions modelling the field prior, like the top one which has a Gaussian form, or likelihood, like the bottom right ones. White ellipses are statistical variables. The two blue ellipses show the input from the data in the inference.   \label{fig:BHM}}
\end{figure}

This section provides a detailed overview of the method used in this work.
We focus on the salient features of the model that we have adopted to analyse the \sdssiii{} data. We note that all the expressions are written as for only one galaxy catalogue. To reduce the number of indices, we have omitted to explicitly mention every-time that the inference is given a set of different independent galaxy catalogue. However the method as implemented does take this into account, and the expressions may be trivially generalised to the multi-catalogue case. We thus omit this in the rest of the section.

In Section~\ref{sec:dm_dynamics}, we give a short presentation of the statistical modelling and sampling algorithms, with references to our previous work where details can be found. Then we present the dynamical model that used in this work in Section~\ref{sec:lpt}, before describing the galaxy predictive model, i.e. the bias model, in Section~\ref{sec:galaxy_bias}. In Section~\ref{sec:foreground_templates}, we present the foreground templates used to model known systematic effects. In Section~\ref{sec:robust_likelihood}, we move on our likelihood, designed to absorb most of the other, unknown, systematic effects. Finally in Section~\ref{sec:sys_map_inference}, we show how this likelihood can be reversed to provide an inference procedure for the systematic maps themselves.

\subsection{A probabilistic physical forward model of dark matter dynamics}
\label{sec:dm_dynamics}

As mentioned in the introduction, this work describes the extension and application of our previously-developed \borg{} algorithm. \borg{} aims at inferring a fully probabilistic and physically plausible model of the three-dimensional matter distribution from observed galaxies in cosmological surveys
\citep[see e.g.][]{jasche_bayesian_2013,jasche2015,lavaux_unmasking_2016}. This framework solves a large-scale Bayesian inverse problem by fitting a dynamical structure formation model to data and inferring the primordial initial conditions from which presently-observed structures formed. The development of this framework was stemmed by convincing evidence from Cosmic Microwave Background observations that the statistics of initial conditions are close to Gaussian \citep{PlanckCollaboration2019_fNL}. In contrast, the statistics of present large-scale structures are very complex and strongly non-Gaussian. It happens that modelling the change of coordinates relating initial conditions to visible large-scale structures is not so complicated in the cosmological paradigm, and it is even feasible to sample the initial states with a Monte Carlo algorithm, currently the Hamiltonian Markov Chain Monte Carlo algorithm \citep{jasche_bayesian_2013}.
This physical forward modelling approach naturally accounts for the formation of non-linear and non-Gaussian large-scale structures, associated with statistics of the density field beyond 2-point correlations, redshift-space distortions and light-cone effects.
As a result, the algorithm provides plausible three-dimensional matter density fields, but also performs a full four-dimensional state inference and recovers the dynamic formation history and velocity fields of the cosmic large-scale structures.

The method also accounts for systematic and stochastic uncertainties, such as survey geometries, selection effects, unknown noise and galaxy biases, as well as foreground contamination \citep[see e.g.][]{jasche_bayesian_2013,jasche2015,lavaux_unmasking_2016,Jasche2017,porqueres_explicit_2019}. For further details on the statistical inference machinery and solutions to the described large scale Bayesian inverse problem, the reader is referred to our previous work \citep[][]{jasche_bayesian_2013,jasche2015,lavaux_unmasking_2016,Jasche2019_PM}.

We note that our model includes many components (initial Fourier modes amplitudes and phases, galaxy bias and foreground contamination). Parameters related to these components are all injected in a probabilistic framework which is bound together on one side by Bayesian priors, typically Gaussian initial conditions, and by the likelihood of galaxy observations on the other side. A graphical summary of all the steps and connections involved in this Bayesian inference is given in Figure~\ref{fig:BHM}. As already detailed many times in our previous work,  we use \emph{sampling} algorithms to build a fair ensemble of points in the posterior parameter space, providing globally a numerical approximation to the posterior density distribution. We do \emph{not use an iterative} procedure which would provide a single answer for the reconstruction problem, and would potentially bias the result as in the case of the Wiener filter \citep{Rybicki1992}. We want to stress this point to reduce misconceptions about our results.

\subsection{Dynamics and light cone model}
\label{sec:lpt}

The target resolution of the inferred initial conditions and modelled galaxy distribution, discussed in the Section~\ref{sec:data}, is about 16\Mpch{}. At this  resolution, the mass density on the mesh at that scale is only affected by mildly non-linear dynamics at low redshift. We thus limit our model of the dynamics in \borg{} to first-order Lagrangian Perturbation Theory (LPT), also called the Zel'dovich approximation \citep{ZelDovich1970,BOUCHET1995}.
 This model predicts the matter density field and its dynamics with sufficient accuracy at the scales relevant to this work \citep[$\sim 16$\Mpch{}, e.g.][]{Bernardeau2002}.

In \borg{}, we use a particle representation to implement the LPT model. The relation between the final Eulerian position $\mvec{x}$ and the initial Lagrangian position $\mvec{q}$ of each  particle is given as:
\begin{equation}
    \mvec{x}(\mvec{q}) = \mvec{q} + D_{+}(a_f) \mvec{\Psi}(\mvec{q}),
\end{equation}
with $\mvec{\Psi}(\mvec{q})$ the displacement field derived from the initial conditions proposed by the sampling algorithm, and $D_{+}(a_f)$ the linear growth factor at the scale factor $a_f$. We encourage the reader to refer to the previous publications \citep{jasche_bayesian_2013,Jasche2019_PM} for further details on the numerical implementation of the forward and adjoint gradient. For the purpose of this work, we further modify the model by adjusting $D_{+}$ depending on the distance to the observer to simulate a light cone. The new evolution equation is thus
\begin{equation}
    \mvec{x}(\mvec{q}) = \mvec{q} + D_{+}(a(|\mvec{q}|)) \mvec{\Psi}(\mvec{q}),
\end{equation}
with $a(d)$ the relation between the comoving distance $d$ and the scale factor of the homogeneous Universe at that look-back distance. This model is an approximation to the full problem of light cone building which involves computing intersections of two trajectories, the trajectory of the light emitting object and the geodesic of the photon emitted by that object and detected through observatories. The average error produced by
this approximation is given by the typical amplitude of the displacement field $\mvec{\Psi}(\mvec{q})$. In a \lcdm{} universe with Planck cosmology \citep{PlanckCollaboration2018}, that displacement is of order 5\Mpch{}, with a maximum of $\sim$20\Mpch{} for the fastest moving objects, which is of the same size as a volume element of our mesh (see Appendix~\ref{app:displacement}). At the farthest distance, this approximation means that we neglect additional coherent distortion of about 10\Mpch{}. At the present resolution, this is still acceptable. However, future improvement on the resolution will require to investigate the detailed impact of the light cone effect on the reconstruction.

\subsection{Galaxy bias model}
\label{sec:galaxy_bias}

In this work, we introduce a new bias model which is built on a few requirements: i/ galaxy formation is a non-local process in Eulerian coordinates, meaning that the model must be somehow sensitive to the environment, and not necessarily linearly; ii/ it should follow features of the linear bias model as much as possible on large scales, for better comparison with earlier literature, and (more interestingly) to offer a connection to generic perturbative bias expansion \citep{Desjacques2018,Schmidt2019,Elsner2019}; iii/ it must ensure positivity of the final galaxy population field to give physically meaningful predictions. As such, we  introduce the following bias model to predict the number of galaxies $N^{(g)}_i$ in a mesh element indexed by $i$:
\begin{equation}
    N^{(g)}_i = \mvec{\Delta}_i^\dagger \mmat{Q} \mvec{\Delta}_i, \label{eq:quadratic}
\end{equation}
with $\mmat{Q}$ a positive definite matrix and $\mvec{\Delta}_i$ a vector formed from local averages of the matter density contrast field $\delta$. In order to guarantee that any sampled matrix is positive definite, we use the Cholesky decomposition of $\mmat{Q}$ as sampling parameters, i.e. the matrix $\mmat{L}$ with $\mmat{Q} = \mmat{L} \mmat{L}^\dagger$. The vector $\mvec{\Delta}_i$ is defined as follows:
\begin{equation}
    \mvec{\Delta}_i^\dagger= \left(1, \delta^{(1)}_i, (\delta^{(1)})^2_i, \ldots, \delta^{(2)}_i, \left(\delta^{(2)}_i\right)^2, \ldots, \left(\delta^{(3)}_i\right), \ldots\right),
\end{equation}
with $\delta^{(\ell)}_i = A^{(\ell)}(\{\delta_i\})$, $A^{(\ell)}$ being an averaging operation in a neighbourhood of the $i$-th mesh element for $\ell \geq 2$ or the identity for $\ell = 1$. This can be written more compactly as
\begin{equation}
    \Delta_{i,a} = \left(\delta^{(\ell_a)}\right)^{\gamma_a}
\end{equation}
for $\ell_a \geq 0$ and $\gamma_a \geq 0$.
In this work, the averaging is done in practice with an oct-tree structure. The level $\ell=1$ is directly the density fluctuation at the finest level, i.e. $\delta$.  For higher levels, $\ell > 1$, we derive the density fluctuations using the following relation:
\begin{equation}
    \delta_{x,y,z}^{(\ell)}(\{\delta_i\} ) = \frac{1}{8^\ell} 
        \sum_{a,b,c=0}^{2^{(\ell)}-1} \delta_{m^{(\ell)}(x,a), m^{(\ell)}(y,b),m^{(\ell)}(z,c)}\;,
    \label{eq:coarsen_average}
\end{equation}
with the coarsening operator
\begin{equation}
    m^{(\ell)}(x,a) = 2^{\ell} \floor{x/2^{\ell}} + a\;. \label{eq:coarsening}
\end{equation}
For the purpose of Hamiltonian Markov Chain exploration used in \borg{}, we compute analytically and provide the adjoint gradient of the above model in Appendix~\ref{app:ag_bias}.

\subsection{Foreground templates}
\label{sec:foreground_templates}

A major point of contention in data analysis is the level of systematic effect contaminating the observational data. The contamination affects the spectroscopic sample of galaxies by hindering a proper uniform target selection from pure photometry. Indeed, to build a galaxy sample, one must generally start with broadband photometry, from which a list of candidates for spectrum measurement is built. Once its spectrum is measured, each candidate object is classified, e.g. as a star or a galaxy. Any bias in target selection can affect the resulting samples of classified objects. For this reason, the final spectroscopic sample of galaxies reflects the biases of the target selection procedure. 

In the case of \sdssiii{}, several groups have studied the possible implication of different contaminants \citep[e.g.][]{ross_clustering_2012,leistedt_exploiting_2014}. In this work, we follow the model presented in \citet{Jasche2017} to represent the effect of a small number of systematics maps, which we use to benchmark the effectiveness of the robust likelihood mechanism presented in the next section. The assumed model is multiplicative, i.e.
\begin{equation}
    N_i^{\text{(g, obs. pred.)}} = R_{i} \prod_{a}(1+\alpha_a F_{a,i}) N_i^\text{(g)}\;, \label{eq:multiplicative_foreground}
\end{equation}
with $N_i^\text{(g, obs. pred.)}$ the predicted mean number of galaxies at mesh element $i$ that is observed given observational constraints (mask and systematics), and $N_i^\text{(g)}$ the predicted mean number of galaxies from the dynamical model in the same mesh element, as obtained in Equation~\eqref{eq:quadratic}. In the above equation, we have also introduced $R_{i}$ the  linear survey response, which accounts for the mask and the  selection effects (radial and angular), $F_{a,i}$ the value of the foreground template $a$ in mesh element $i$, and $\alpha_a$ the  intensity of foreground $a$. The linear response is generally provided as part of the meta-data of a given survey. It is estimated from the target and spectroscopic sample. In the case of \sdssiii{}, that is just the ratio between those two samples for each angular direction. The parameter $\alpha_a$ is left free  sampled directly from its posterior given the data. As mentioned in the introduction of Section~\ref{sec:method}, there is one parameter for each foreground and for each for \emph{each catalogue} part of the inference problem.
This multiplicative foreground model can reasonably model a broad class of systematic effects, such as intergalactic absorption of light by dust, atmospheric effects or fibre collisions. However, it is limited to known effects for which sky models exist.

\subsection{Robust likelihood}
\label{sec:robust_likelihood}

The \sdssiii{} survey has been designed to optimally study galaxy clustering at the scales of BAOs. While control of systematic effects at these scales has been studied in detail by the SDSS collaboration \citep[e.g.][]{Reid2015}, there exists no equally-good understanding of the impact of systematic effects at the largest scales of the galaxy distribution, typically for modes of wavenumber $k \la 10^{-2}h$~Mpc$^{-1}$. So far, state-of-the-art data analysis methods have had limited success in removing some of the large-scale systematic effects inherent to the observations \citep{kalus_map-based_2019}. These results indicate that there probably exists a scale in data beyond which galaxy clustering is not understood because it is not modelled sufficiently well using known foreground templates. To address this issue, \citet{porqueres_explicit_2019} developed a new likelihood, based on Poisson statistics, which is designed to be robust against unknown foreground contamination at a scale given \textit{a priori}. The underlying idea relies on the assumption that the physically modelled galaxy distribution can be related to the observed one up to some overall scaling over patches on which the unknown foreground modulation is quasi constant. These patches can be chosen in any convenient way for the analysis. In practice, we use a three-dimensional extrusion of pixels of a \healpix{} map, yielding a 3d patch map. This allows us to group pixels with quasi constant foreground amplitudes into sets
$\mathcal{A}_m$, where $m$ runs over indices of an \healpix{} map. The effective predicted Poisson intensity is $A_{p_i} N^\text{(obs. pred.)}_i$, where $i$ is a mesh element index of the 3d grid covering the considered volume of Universe, $p_i$ the index of the patches containing $i$ (otherwise said $i \in \mathcal{A}_{p_i}$), and $N_i^\text{(g, obs.)}$ the raw galaxy count intensity predicted by the dynamical model and the galaxy bias model. We build the following probabilistic model:
\begin{multline}
    P(\{N^\text{(g, obs.)}_i\}, \{A_p\} | \{N_i^\text{(obs. pred.)}\}) = \\
    \prod_{p} \left[\prod_{i \in \mathcal{A}_p} \text{Poisson}(A_{p} N^\text{(g, obs. pred.)}_i)\right] \pi(A_p)\;, \label{eq:full_joint_robust}
\end{multline}
with $\{N^\text{(g, obs.)}_i\}$ the number count of galaxies observed in the voxel $i$.
We choose $\pi(A) \propto 1/A$ as the prior probability of the amplitude of the unknown systematic, which thus follow a Jeffreys' prior. The problem of large scale structure inference uses the marginalised version of that probability, and was shown to be resilient to unknown systematics in a test on mock data \citep{porqueres_explicit_2019}. After marginalisation over $\{ A_p \}$, the new likelihood takes a simple form:
\begin{multline}
    P\left(\{N^\text{(g, obs.)}_i\} | \{ N^\text{(obs. pred.)}_i \}\right) \propto \\
    \prod_{p} \prod_{i\in \mathcal{A}_p} \left(\frac{N^\text{(obs. pred.)}_i}{\sum_{j \in \mathcal{A}_p} N^\text{(obs. pred.)}_j} \right)^{N^\text{(g, obs.)}_i}\;.
\end{multline}
We immediately notice that this likelihood is insensitive to absolute scales in the predicted galaxy number intensity $\{N^\text{(obs. pred.)}_i\}$, which is an appealing feature: the ratio cancels any contribution over a scale corresponding to the assumed smoothness of the foreground contamination. Robustness tests are described in more details in \citet[]{porqueres_explicit_2019}.

\begin{figure*}
    \centering
    \includegraphics[width=\hsize]{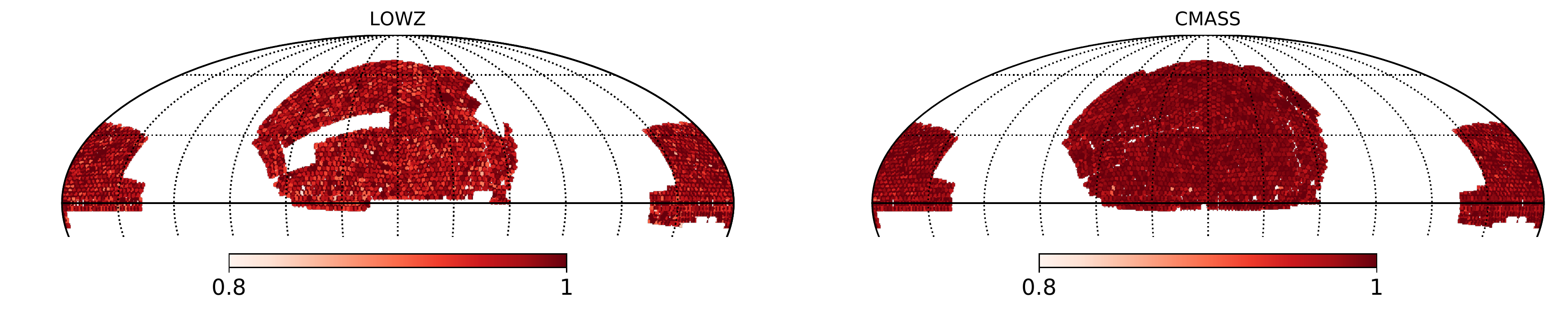}
    \caption{Completeness maps of \sdssiii{} for the LOWZ sample (left panel) and CMASS sample (right panel). These completeness maps are directly derived from the DR12 repository and rendered on an \healpix{} mesh at $N_\text{side}=2048$. We note the usual vetoed regions in LOWZ corresponding to a problem in the target selection that occurred during the first year of data acquisition  \citep{Parejko2013}.}
    \label{fig:boss_completeness}
\end{figure*}

\begin{figure*}
    \includegraphics[width=\hsize]{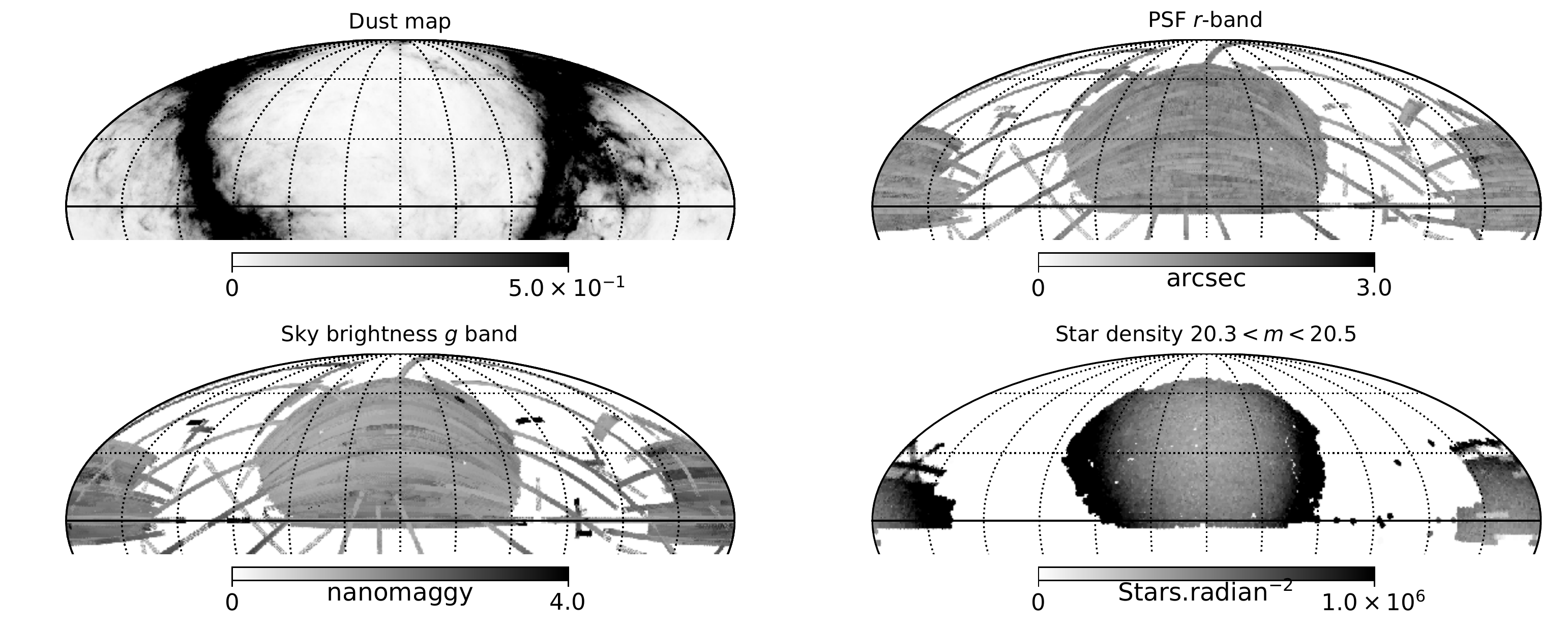}
    \caption{Four of the eleven known systematic maps that we have used in the inference presented in this work. Each of these maps was rendered at $N_\text{side}=2048$ from a {\sc mangle} representation. The above are dust induced reddening (\textit{top left}), point spread function (PSF) in the $r$ band (\textit{top right}), sky flux in the $g$ band (\textit{bottom left}), density of stars with an apparent magnitude $20.3<i<20.5$ (\textit{bottom right}). We note the typical striping induced by the SDSS scanning strategy for the top right and bottom left maps.}
    \label{fig:systematic_map_prior}
\end{figure*}

\subsection{Systematic map inference}
\label{sec:sys_map_inference}

The robust likelihood is designed to ignore information on some spatial scale. In the case of this work, we limit ourselves to ignore information above some angular scale, even if the framework would also work also for complex 3d scales. However, we may still use the inferred density field to solve the reverse problem of inferring the systematic effects that were ignored within the Markov Chain Monte Carlo analysis (MCMC, see Section~\ref{sec:results}). In doing so, we obtain complete maps of the unknown systematic effects down to some angular scale. For one patch $\mathcal{A}_p$, we may derive the conditional probability of the value taken by $A_p$ from Equation~\eqref{eq:full_joint_robust}
\begin{multline}
    P(A_p | \{N^\text{(g, obs.)}_i\}, \{N^\text{(obs. pred.)}_i\}) \propto \\
    \frac{1}{A_p} \prod_{j \in \mathcal{A}_p} \text{Poisson}\left(A_{p} N^\text{(obs. pred.)}_i\right)\; .
\end{multline}
For most purposes, we are only interested in the first two moments of the above distribution, the mean and the variance. These may be computed analytically:
\begin{align}
    \mu_{A_p|\mvec{N}^\text{(obs. pred.)}} & = \langle A_p \rangle = \frac{\sum_{i\in \mathcal{A}_p} N^\text{(obs)}_i}{\sum_{i \in \mathcal{A}_p} N^\text{(obs. pred.)}_i} \;, \label{eq:mean_robust_map} \\
    \sigma^2_{A_p|\mvec{N}^\text{(obs. pred.)}} &= \langle (A_p-\mu_{A_p})^2 \rangle \nonumber \\
    &= \frac{\sum_{i\in \mathcal{A}_p} N^\text{(obs)}_i}{\left(\sum_{i \in \mathcal{A}_p} N^\text{(obs. pred.)}_i \right)^2}\;. \label{eq:variance_robust_map}
\end{align}
In the above, we have used the following identity to compute the integral over the Poisson distribution:
\begin{equation}
    \int_0^{+\infty} \text{d}x\; x^\alpha \exp(-\beta x) = \alpha! \beta^{-\alpha-1}\;.
\end{equation}
We note that we did not specify the derivation of the set $\mathcal{A}_p$ for each patch of interest. In our case, we are interested in computing sky maps at different redshift of detectable systematic effects, given our model of large scale structures. We use the \healpix{} pixelization to represent these maps. Each patch thus corresponds to the cosmological volume that projects in each pixel of the sought map. We build the set $\mathcal{A}_p$ by throwing 100 uniformly distributed rays at random within each pixel and recording the voxels that are traversed. 
Because voxel has a finite size, many rays for different \healpix{} pixels will traverse the same voxels. This means that the maps derived from this procedure will have nearby pixels with highly correlated values. That is not a fundamental limitation but a choice of representation of the systematic map that we aim to derive. The value for the patches derived from the posterior analysis would be completely decorrelated if we had decided to choose a non-overlapping set of voxels to compute pixel values.

We note that the average and the variance per pixel given in Equations~\eqref{eq:mean_robust_map} and \eqref{eq:variance_robust_map} are for one particular model of large-scale structure given by the set of values $\left\{N^\text{(g, obs. pred.)}_i\right\}$. \borg{} provides an ensemble of plausible values for this field. The probability for the set of pixels $\{A_p\}$ is thus:
\begin{multline}
    P(\{A_p\} | \{N^\text{(g, obs.)}_i\}) = \\
     \int \left(\prod_{j=1}^{N_\text{g}}\text{d}N^\text{(obs. pred.)}_j \right)
     P(\{N^\text{(obs. pred.)}_j\}|\{N^\text{(obs)}_i\})\; \times \\
     P(\{A_p\} | \{N^\text{(obs)}_i\}, \{N^\text{(obs. pred.)}_j\})  \\
     \simeq \frac{1}{N_\mathrm{sample}}\sum_c P(\{A_p\} | \{N_i\}, \{N^\text{(obs. pred.)}_{i,c}\})\;,
\end{multline}
with $N^\text{(obs. pred.)}_{i,c}$ the predicted observed galaxy intensity in voxel $i$ for the Markov chain sample $c$, $N_\mathrm{sample}$ the number of considered samples in the MCMC, and $N_g$ the number of mesh element to represent the matter density field.
Thus, the marginalised mean and variance at each pixel position is computed by taking the average over the Markov chain of the mean and variance given by Equations~\eqref{eq:mean_robust_map} and \eqref{eq:variance_robust_map}.

\begin{figure*}
    {
        \centering
        \includegraphics[width=\hsize]{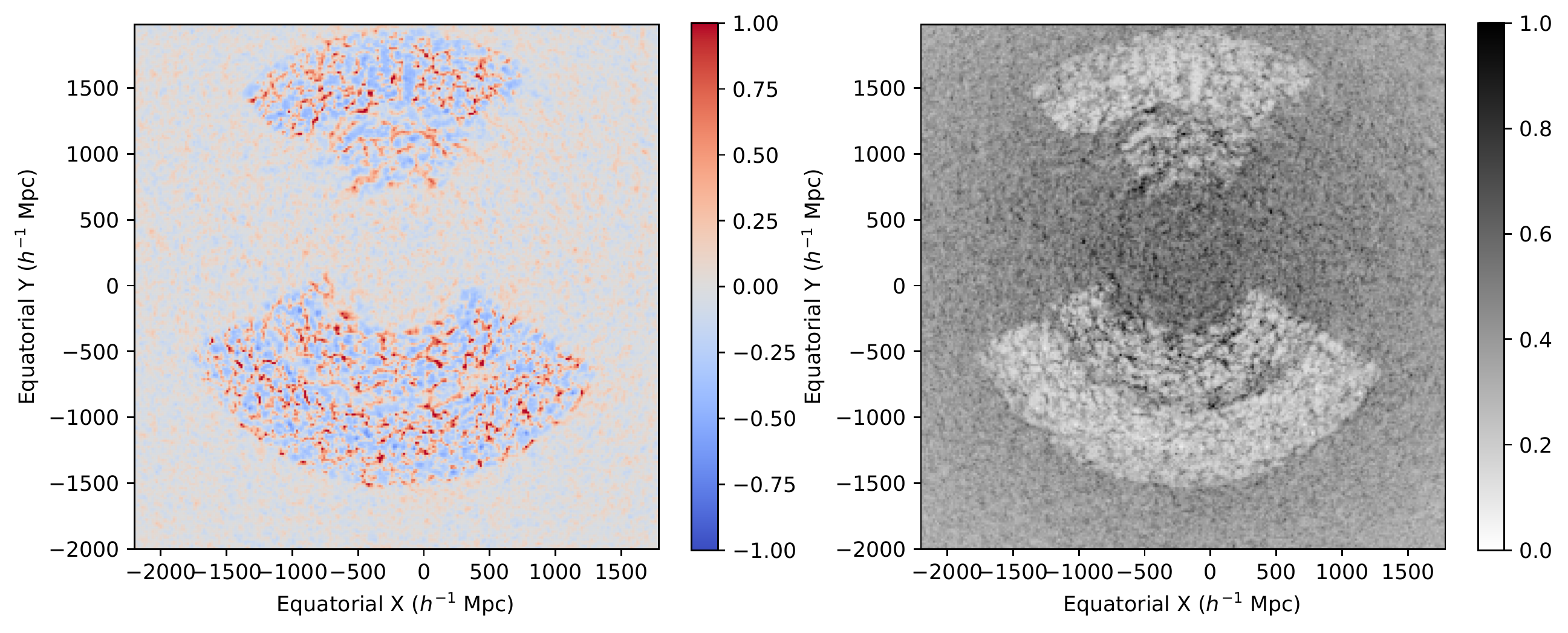}
    }
    \caption{Slices through the inferred three-dimensional ensemble mean density field (\textit{left panel}) and the corresponding standard deviations of density amplitudes (\textit{right panel}) obtained from the Markov Chain. As can be seen, the algorithm recovers the filamentary large-scale structures in regions sampled by galaxies of the \sdssiii{} survey, while matter density approaches cosmic mean in unobserved regions. Correspondingly, the map of standard deviations shows low variance in observed regions and correctly provides higher uncertainty in unobserved regions. The plot illustrates that \borg{} provides detailed reconstructions of matter density fields and corresponding uncertainty quantification. The coordinates are all comoving assuming the cosmology given in Section~\ref{sec:results}. The residual small fluctuations outside the observed region in the left panel are due to the finite length of the Markov Chain.}
    \label{fig:density}
\end{figure*}

\begin{figure*}
    \centering
    \includegraphics[width=\hsize]{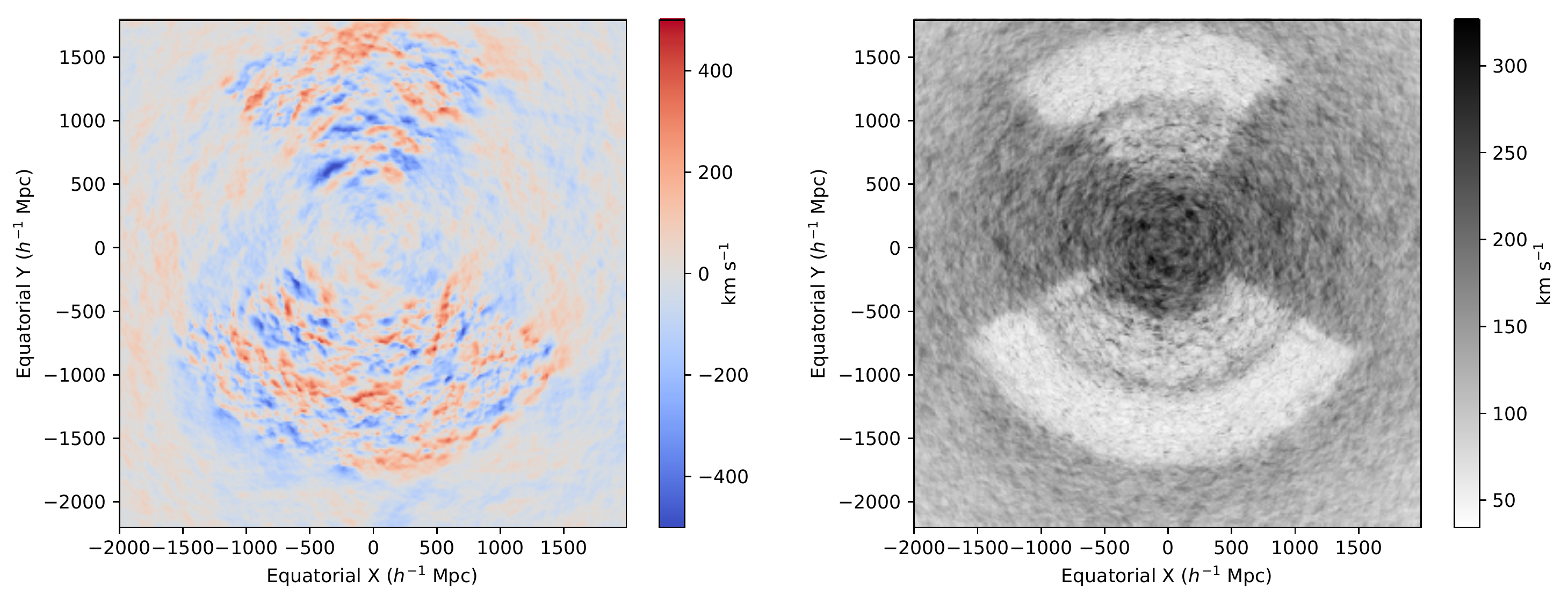}
    \caption{Slices through the cosmic velocity field derived with \borg{} using data from \sdssiii{}.\textit{Left panel}: ensemble average of the line-of-sight component of the velocity field. \textit{Right panel}: standard error of the mean of the velocity field (standard deviation derived from the posterior). The effect of the light cone is visible (giving a factor of $\sim$2 difference in the standard deviation between $z=0$ and $z=0.7$).}
    \label{fig:velocity}
\end{figure*}

\section{The \sdssiii{} data}
\label{sec:data}

\begin{table}
    \centering
    \begin{tabular}{cp{.7\hsize}}
        \hline
         Name & \multicolumn{1}{c}{Definition} \\
         \hline
         dust & Dust induced reddening \citep{SFD} \\
         sky flux & Photometric sky flux in the indicated \newline band  (5 maps) \\
         airmass $r$ & Air mass above telescope, $r$ band \\
         psf $r$ & Point spread function, $r$ band \\
         star 0 & Density of stars with $20.5 < i < 20.3$ \\
         star 1 & Density of stars with $20.3 < i < 20.1$  \\
         star 2 & Density of stars with $20.1 < i < 19.0$ \\
         \hline
    \end{tabular}
    \caption{Name convention for the 11 foreground templates used in this work. The coefficients $\alpha_a$ attached to each of these templates in Equation~\eqref{eq:multiplicative_foreground} are sampled jointly with the bias parameters and the matter density field.}
    \label{tab:systematics}
\end{table}

We apply our Bayesian inference approach to galaxies observed by the Baryon Oscillation Spectroscopic Survey \citep[BOSS,][]{Dawson2013}, the third generation of the Sloan Digital Sky Survey \citep[SDSS-III,][]{Einstein2011_SDSS3}.
The BOSS survey is dedicated to observing the three-dimensional clustering of $1.37$ million galaxies with spectroscopic redshifts covering about $10\;000$~$\mathrm{deg}^2$ of the sky over two contiguous regions in the Northern and Southern Galactic caps. This work uses the final data release DR12\footnote{\url{https://data.sdss.org/sas/dr12/boss/lss/}} of \sdssiii, containing the data of all six years of the survey \citep[][]{alam_eleventh_2015}. Galaxies were targeted uniformly in a low-redshift sample with $z<0.45$ (LOWZ). To select additional massive galaxies in the redshift range $0.4 < z < 0.8$, several colour cuts were applied to the SDSS-III imaging data in the ($u$,$g$,$r$,$i$,$z$) bands \citep[][]{fukugita_sloan_1996}. According to the passively evolving model, these selections result in a sample (CMASS) that has a constant stellar mass limit over the redshift range $0.4 < z < 0.8$ \citep[][]{maraston_modeling_2009}. Large stellar masses imply strong galaxy biases with respect to the underlying dark matter density field. This property induces that each galaxy provides strong indication of large scale matter fluctuations, yielding more information on large scale structure analysis than their lower stellar mass counterpart.
Detailed descriptions of BOSS targeting criteria, data reduction methods and the construction of the large-scale structure catalogue are described in \citet{Einstein2011_SDSS3,Dawson2013,alam_eleventh_2015,Reid2015}.
More specifically, we use large-scale structure catalogues provided by the BOSS galaxy clustering working group \citep[][]{Anderson2014,Reid2015}. Their sample assigns weights to galaxies to correct for non-cosmological fluctuations imprinted on the target catalogue by imperfections in the acquisition of spectroscopic redshifts due to fibre collisions, precluding simultaneous assignments of spectroscopic fibres to targets closer than $\ang{;;62}$ \citep[][]{Ross11,Ho2012,ross_clustering_2012}. Additional weights are assigned to compensate systematic effects between observed galaxy number densities and the seeing \citep[][]{Reid2015}. To account for survey geometry and spectroscopic completeness, we used the {\sc mangle} software \citep[][]{SWANSON2008MNRAS} to create corresponding \healpix{} \citep[][]{HEALPIX} maps at $N_{\mathrm{side}}=2\,048$ shown in Fig.~\ref{fig:boss_completeness}.
To account for redshift dependence in radial selection functions we split the LOWZ and CMASS sample into four redshift bins and two galactic caps each, making a total 16 sub-catalogues. We have thus 4 redshift bins for each of the following catalogue: LOWZ north galactic cap (NGC), LOWZ south galactic cap (SGC), CMASS NGC and CMASS SGC.
The four distance bins for LOWZ are chosen as  [600, 750]\Mpch, [750, 900]\Mpch, [900, 1050]\Mpch, [1050, 1200]\Mpch. Similarly we have four sub-catalogues for CMASS with distance bins chosen as [1000, 1200]\Mpch{}, [1200, 1400]\Mpch{}, [1400, 1600]\Mpch{}, [1600,1800]\Mpch{}.

The \borg{} algorithm permits us to treat the respective systematic effects, such as redshift dependent galaxy biases, selection effects and foreground contamination, for each of these 16 galaxy sub-samples. In the following we will provide more detail to our data analysis procedure.

While not strictly required by our robust likelihood framework, we also derive systematic maps from the meta-data of the \sdssiii{} photometric database. As mentioned in Section~\ref{sec:foreground_templates}, we make them part of the data model to learn about specific features that could still be there even after the cleaning performed by the robust likelihood. We have included 11 foregrounds which have been classically considered by the \sdssiii{} collaboration \citep[e.g.][]{ross_clustering_2012} and which still potentially contaminate the data, despite applying weights to galaxies. The inferred amplitude values will be used to assess the amount of residual corrections that are still necessary at small scales. The known systematic effects that we consider are summarised in the Table~\ref{tab:systematics}. We note that we consider the impact of the sky flux independently in each of the SDSS photometric band ($u$,$g$,$r$,$i$,$z$), thus the line ``sky flux'' corresponds actually to five maps. Each of these maps was derived using the {\sc mangle} \citep{SWANSON2008MNRAS} geometry file describing the \sdssiii{} large structure sample.\footnote{\url{https://data.sdss.org/sas/dr12/boss/lss/geometry/boss_geometry_2014_05_28.fits}} Each {\sc mangle} polygon was assigned a weight depending on the meta data of the corresponding photometric tile. We rendered the maps on an \healpix{} mesh at $N_\text{side}=2\,048$. This corresponds to a precision of $\sim$2~arcminutes. As an illustration, we show a subset of four of the eleven foregrounds maps in Figure~\ref{fig:systematic_map_prior}. These maps show, in Mollweide projection, the vector $F_{a,i}$ of Equation~\eqref{eq:multiplicative_foreground}.

\begin{figure}
    {
        \centering
        \includegraphics[width=\hsize]{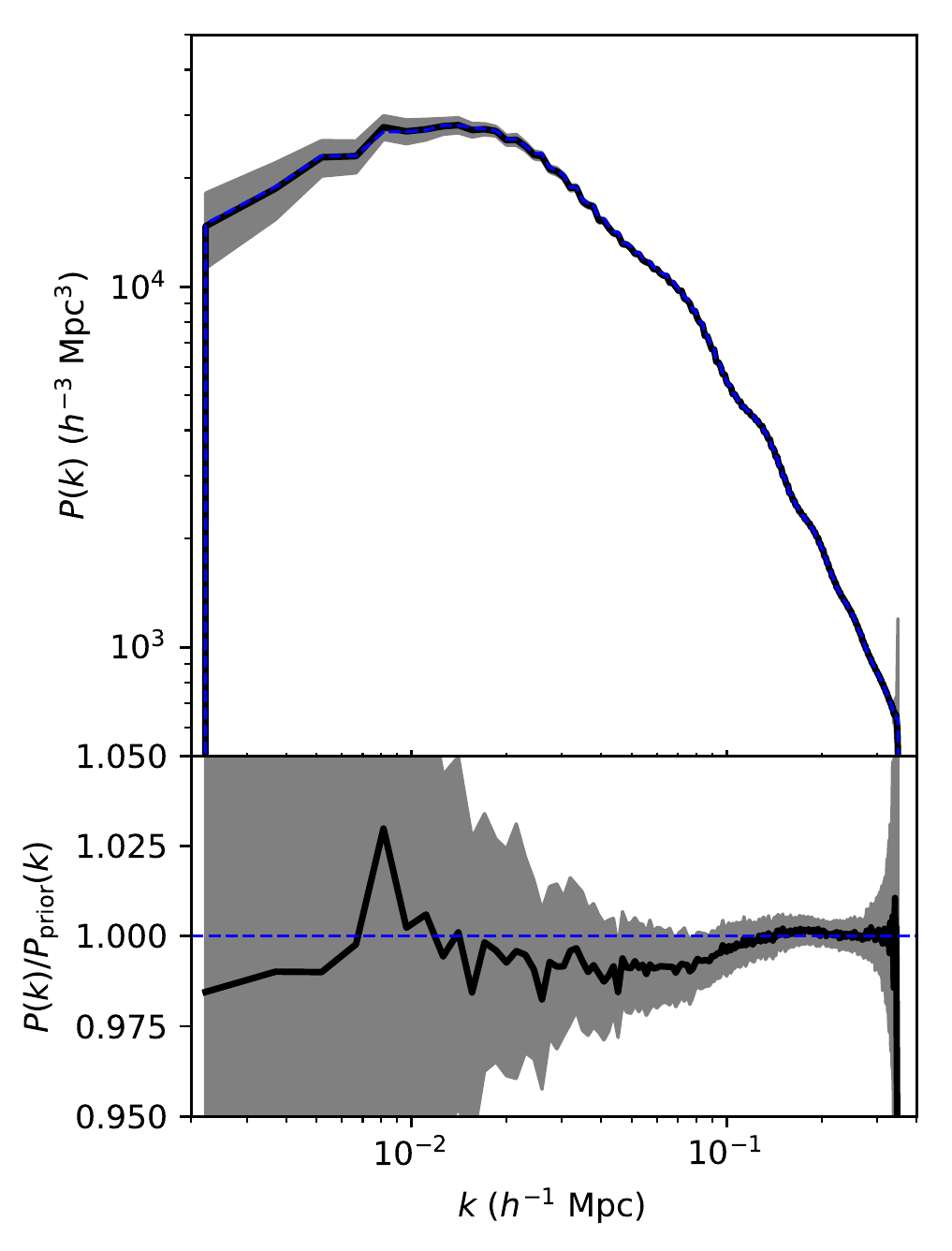}
    }
    \caption{Ensemble average power spectrum of the \textit{a posteriori} samples of initial conditions produced by \borg{}. \textit{Top panel}: Mean of the power spectra (black line), one standard deviation (shaded grey region), and the prior power spectrum (dashed blue line).  \textit{Bottom panel}: same as the top panel but we divide by the prior power spectrum before plotting. The finest mode reachable in the box on the right-hand side is at $k\simeq 0.35$\hMpc{} as can be seen by the strong increase of variance at that scale. It corresponds to a scale that already includes non-linear features. Deviations on large scales, up to the largest mode at $k \simeq 10^{-3}$\hMpc{}, are below a few percents, confirming the systematic-free nature of the density reconstruction. }
    \label{fig:final_Pk}
\end{figure}

\section{Results}
\label{sec:results}

In this section, we detail various aspects of our results. In all the following, we have assumed a cosmology close to what the Planck collaboration has found with CMB data \citep{PlanckCollaboration2018}, namely $\Omega_\text{r} = 0$, $\Omega_K=0$, $\Omega_\text{M} = 0.2889$, $\Omega_\text{b}=0.048597$, $\Omega_\Lambda = 0.7111$, $w=-1$, $n_S=0.9667$, $\sigma_8=0.8159$, $H_0 = 67.74$\kmsMpc. We note that the absolute value of the Hubble constant enters only through the transfer function in the power spectrum and does not have an impact on the coordinate transform for example. Early tests indicated that a slightly lower $\Omega_\text{m}$ seems to be preferred. We have thus pushed $\Omega_\text{m}$ to the lower acceptable limit and proceeded with the run. We have used an inference box of 4~000\Mpch{} sampled with a $256^3$ mesh grid, setting resolution to $\sim 15$\Mpch{}. The radial completeness is estimated in a similar way to \citet{Anderson2012}.

The \borg{} inference machine is left free to choose the value of the 14 parameters of the bias model (model described in Section~\ref{sec:galaxy_bias}) and the amplitude of the 11 foreground templates, for each sub-catalogue (Section~\ref{sec:inferred_systematic} and \ref{sec:data} for the description of sub-catalogues), and the amplitude of the $256^3$ modes in the initial condition at $z=1\,000$. We use a particle set of $512^3$ to trace the evolution of matter with Lagrangian perturbation theory. We include both the light cone model of Section~\ref{sec:lpt} and the model of redshift-space distortion effects at the level of particles.
The bias parameters were given a prior centred on zero with a variance unity on elements of $\mmat{L}$, in order to avoid the chain to explore too wide a parameter space initially and cause numerical issues. We have checked that this choice does not impact our result by verifying the \textit{a posteriori} information content: the standard deviation of the sampled parameters are all less 0.7 which is less than 1. 
We have generated $\sim 10~400$ MCMC samples from the posterior distribution. By an analysis of the \textit{a posteriori} power spectrum of the initial density (Appendix~\ref{app:warmup}), we are confident that the chain has reasonably burned in after $\sim$2~000 samples. We use all the samples with an identifier greater than 2~000 for the results shown in this section.

\subsection{Inferred 3D matter density fields}
\label{sec:inferred_density}

The main purpose of our \borg{} inference machine is to derive a probable, dynamical, physical, model of the matter distribution of the observed universe. As such, the first data product that we investigate is the inferred matter density that is required to explain the data. This comes in two forms: the initial conditions, post-recombination but still in the linear regime of the dynamics; and the evolved matter density at the moment the photon is detectable in our light cone. \borg{} samples all possible realisations of the initial conditions that satisfy the observational constraints. We focus our analysis mostly on the first moment of the posterior distribution for each considered parameter. However, more information is available on the posterior.
In Figure~\ref{fig:density}, we show the ensemble average of all realisations of the evolved matter density. We represent a plane parallel to and close to the celestial equatorial plane (DEC=\ang{0;;}), chosen to include the full shape of the light cone probed by the \sdssiii{}. The full density field cube of Figure~\ref{fig:density} will be made available on Zenodo at publication time. We recognise the typical structure of SDSS data: the south galactic cap part (SGC) at $Y>0$, and the north galactic cap (NGC) at $Y<0$. We see a separation between the LOWZ and CMASS components of \sdssiii{} at a comoving distance $\sim 900$\Mpch{}. The distinction is more pronounced in the right panel, showing the standard deviation of fluctuations compared to the mean field: LOWZ clearly yields a noisier  estimate of the matter density than CMASS. Despite the low resolution of our run ($\sim$15\Mpch{} per voxel-side), we see that the density field is non-Gaussian, with some filamentary structure.

In Figure~\ref{fig:velocity}, we show the line-of-sight component of the velocity field, which we infer from the data and the dynamical model. The velocity field is derived using the simplex-in-cell estimator presented in \citet{Abel2012,Hahn2014_DMSHEET,Leclercq2017a} (based on the Delaunay tessellation of elementary Lagrangian cubes of particles into six tetrahedra), applied to the particle tracers generated by \borg{} for each element of the MCMC. We show the ensemble mean average (left panel) and the standard deviation with respect to the mean (right panel). We note that the light cone model also induces a modulation depending from the distance to the observer, located at the centre of the figure. In the right panel for example, the standard deviation of unobserved region is significantly higher close to the observer than in outer regions, as expected in perturbation theory, which predicts a factor $\sim$2 difference between $z=0.2$ and $z=0.7$. Visual inspection of the maps does not show any particular anomaly. We note a slight excess of infall towards the observer in the lower-left region of the left panel, which could be due to an insufficient number of samples in the ensemble averaging of the Markov Chain.

\begin{figure}
    \centering
    \includegraphics[width=.95\hsize]{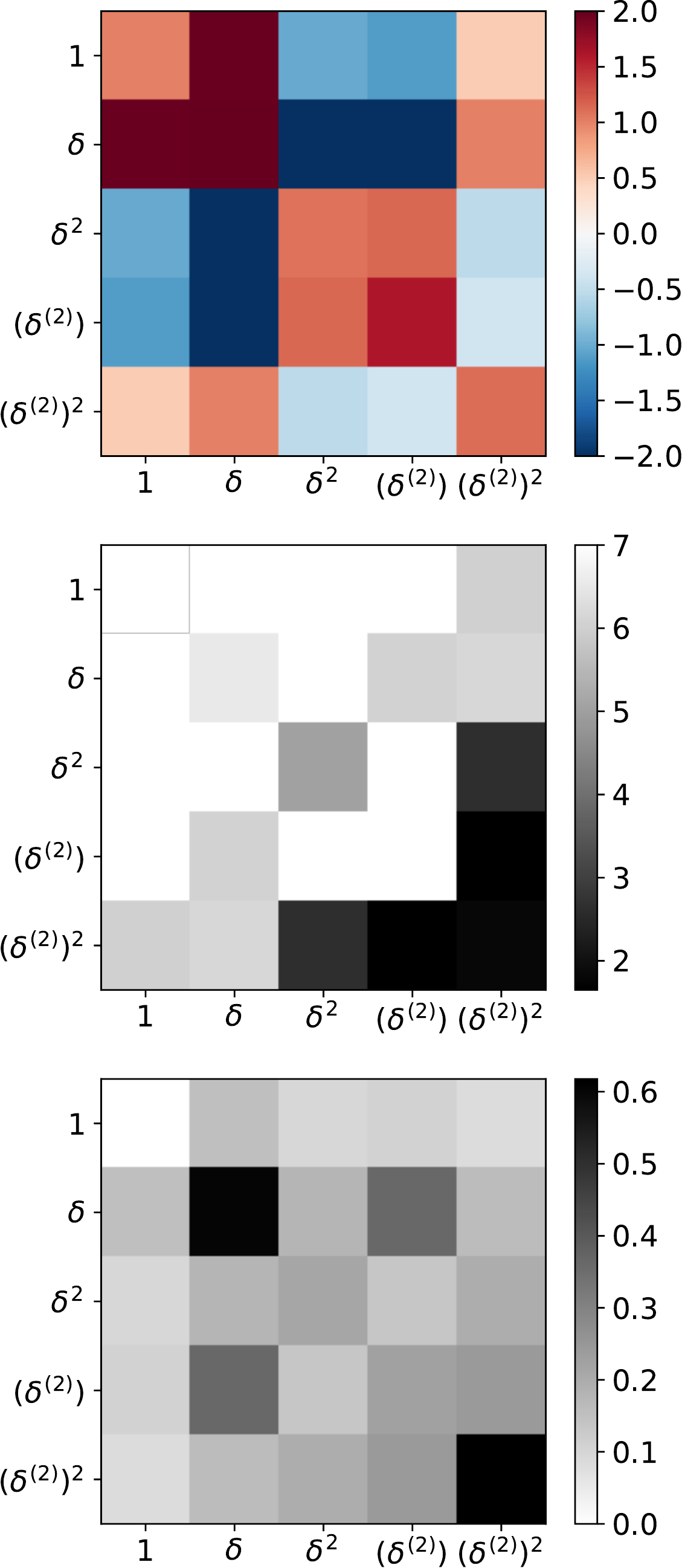}
    \caption{Bias parameters inferred for the CMASS NGC lowest redshift bin. The parameters are presented as the matrix $\mmat{Q}$ appearing in Equation~\eqref{eq:quadratic}. \textit{Top panel}: matrix of mean parameters. \textit{Middle panel}: ratio of the top matrix to the bottom matrix, showing the signal-to-noise matrix for each coefficient. \textit{Bottom panel}: matrix of standard deviations for each coefficient. The inferred matrix represented here corresponds to a quadratic form: each coloured coefficient is associated with the product of the symbols indicated in the corresponding row and column in the bias model (see Equation~\eqref{eq:quadratic}). The off-diagonal terms contribute twice while the on-diagonal terms contribute only once. The coefficient corresponding to $1\times 1$ (top-left corner) is special as it is fixed to a value of unity and not sampled.}
    \label{fig:bias_matrix}
\end{figure}

We show in the top panel of Figure~\ref{fig:final_Pk} the power spectrum of \textit{a posteriori} initial conditions, both the ensemble average and the standard deviation. We also plot our prior on the cosmological power spectrum, obtained from the \citet{Eisenstein1999} fitting function for our choice of cosmological parameters. In the bottom panel, we show the deviations of the \textit{a posteriori} power spectrum from our prior. The prior is not strictly enforced in our inference framework, but only used as a guideline in the absence of informative data. We note that, contrary to previous attempts \citep[e.g.][]{ross_clustering_2012,kalus_map-based_2019}, we do not observe strong deviations of the power spectrum at $k\lesssim 10^{-2}h$~Mpc$^{-1}$. Such deviations are typical of cases where systematic effects are improperly accounted for \citep{porqueres_explicit_2019}. 
If our model did not include systematic effects, the \textit{a posteriori} power spectrum would have received contributions from them at the largest scales, as was noted by all previous attempts as well as our own investigations \citep[e.g.][]{Jasche2017,kalus_map-based_2019,porqueres_explicit_2019}. Given that past Cosmic Microwave Background missions \citep{Bennett2013_WMAP,PlanckCollaboration2018} have not observed any anomalous power at large angular scales deviating from the predictions of the $\Lambda{}$CDM model, we conclude the power spectrum deviations originally observed in \sdssiii{} were due to systematic effects. The fact that we these deviations vanish when using our framework for known and unknown systematic effects indicates that the presence of excess large-scale power is unnecessary to explain data. Furthermore, the posterior captures more information than what is in the prior, which is indicated by the shifted mean in the bottom panel of Figure~\ref{fig:final_Pk}.

\subsection{The galaxy bias model}

The \borg{} forward model includes a new quadratic-form bias, described in Section~\ref{sec:galaxy_bias}. As it is multi-scale, a direct comparison to previous analyses is not straightforward. In Figure~\ref{fig:bias_matrix}, we give an example of the bias parameters that we have inferred, with corresponding uncertainties. These parameters correspond to the coefficients forming the $\mmat{Q}$ matrix for the sub-catalogue holding the CMASS NGC in a distance bin of [1000,1200]\Mpch. This is thus the closest to a central slice in \sdssiii{}. The matrix shown in Figure~\ref{fig:bias_matrix} gives the coefficient of the field produced by the product of its row and column labels in the bias model. For example, coefficients in the top row correspond to coupling the constant (``1'') with something else. If we choose the second column (labelled ``$\delta$''), we obtain the coefficient in the quadratic form that corresponds to the term $1\times \delta$. In the second row and the last column, we find the coefficient of the term $\delta \times \left(\delta^{(2)}\right)^2$. It is a quadratic form, thus off-diagonal terms shall be counted twice owed to the symmetry. In Figure~\ref{fig:bias_matrix}, the top panel is the ensemble average for each coefficient, the bottom panel is the standard deviation with respect to that mean. The middle panel is the ratio of the top to the bottom panel, corresponding to the signal-to-noise ratio.

These bias parameters indicate that there is evidence for scale dependence, as the data require the model to have a non-vanishing second level (terms in $\delta^{(2)}$) to be represented fairly. We also note that it requires some compensation between scales as, for example, in the top row of the top panel of Figure~\ref{fig:bias_matrix}. There, we have a positive coefficient in the second column (``$\delta$'') followed by a  negative coefficient in the fourth column (``$\delta^{(2)}$''). There is thus some evidence of scale-dependent biasing. We leave further interpretation to future work.

\begin{figure}
    \centering
    \includegraphics[width=\hsize]{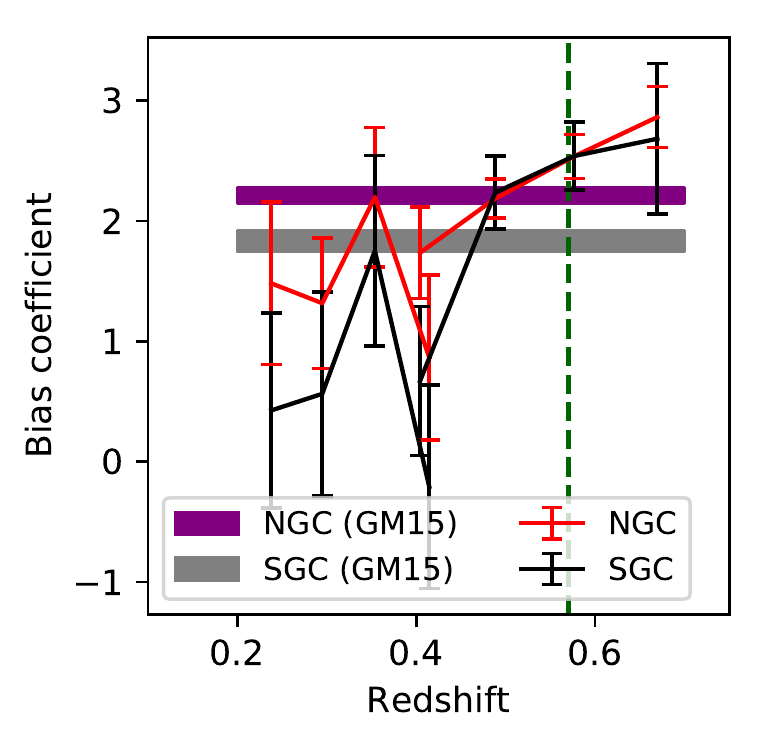}
    \caption{Evolution of the linear component of the bias model with redshift from the two hemispheres. We show here the quantity closest to linear biasing in our model, the sum $2(Q_{1,\delta} +  Q_{1,\delta^{(2)}})$, as discussed in the text. In CMASS (at redshifts $0.4 \leq z \leq 0.7$), the evolution shows evolution with redshift, with a clear increase in the CMASS redshift range. There is little residual difference between data from the northern and southern hemisphere. The measurements of the linear bias coefficient $b_1$ from \citet[GM15]{Gil-Marin2015} are shown by purple and grey bands, with the reference redshift indicated by a vertical dark green dashed line.}
    \label{fig:linear_bias_term}
\end{figure}

As our model is non-local and non-linear (Equation~\ref{eq:quadratic}), it is not straightforwardly related to classical linear biasing. However, assuming that the dark matter density is mostly the same at the two levels $(1)$ and $(2)$, the model has a linear term linking the matter density to galaxy number count which behaves like $2 (Q_{1,\delta} + Q_{1,\delta^{(2)}})$. In Figure~\ref{fig:linear_bias_term}, we show the evolution of this combination of the bias parameters of our model
for the different sub-catalogues, organised in redshift bins. In LOWZ, there is no clear trend for this combination of bias parameters with redshift, but in CMASS, we observe that the equivalent of the linear bias evolves by nearly a factor of two between redshift 0.4 and 0.7. Additionally, there is no big discrepancy between the NGC and SGC part of CMASS, which indicates that the systematic effects between these two regions of the sky have likely been taken care of.

We can only partially compare our results to \citet{Gil-Marin2015} who provide the bias in very large bins, and without light cone correction. As we work with a fixed cosmological prior, we set their $\sigma_8$ to their reference maximum likelihood measurement and focus on the parameter $b_1$. They report their measurement at an effective redshift $z_\text{eff}=0.57$ (indicated by a vertical dashed line), corresponding to our second before last bin in Figure~\ref{fig:linear_bias_term}. Their reported $b_1$ value are indicated by two horizontal bands (purple for NGC, grey for SGC), which corresponds to the measurements reported in table~1 and 2 of \citet{Gil-Marin2015}. Given the respective approximations involved, both measurements agree very well.

\subsection{Cross analysis with CMB lensing}
\label{sec:cross_analysis_cmb}

\begin{figure}
    \centering
    \includegraphics[width=\hsize]{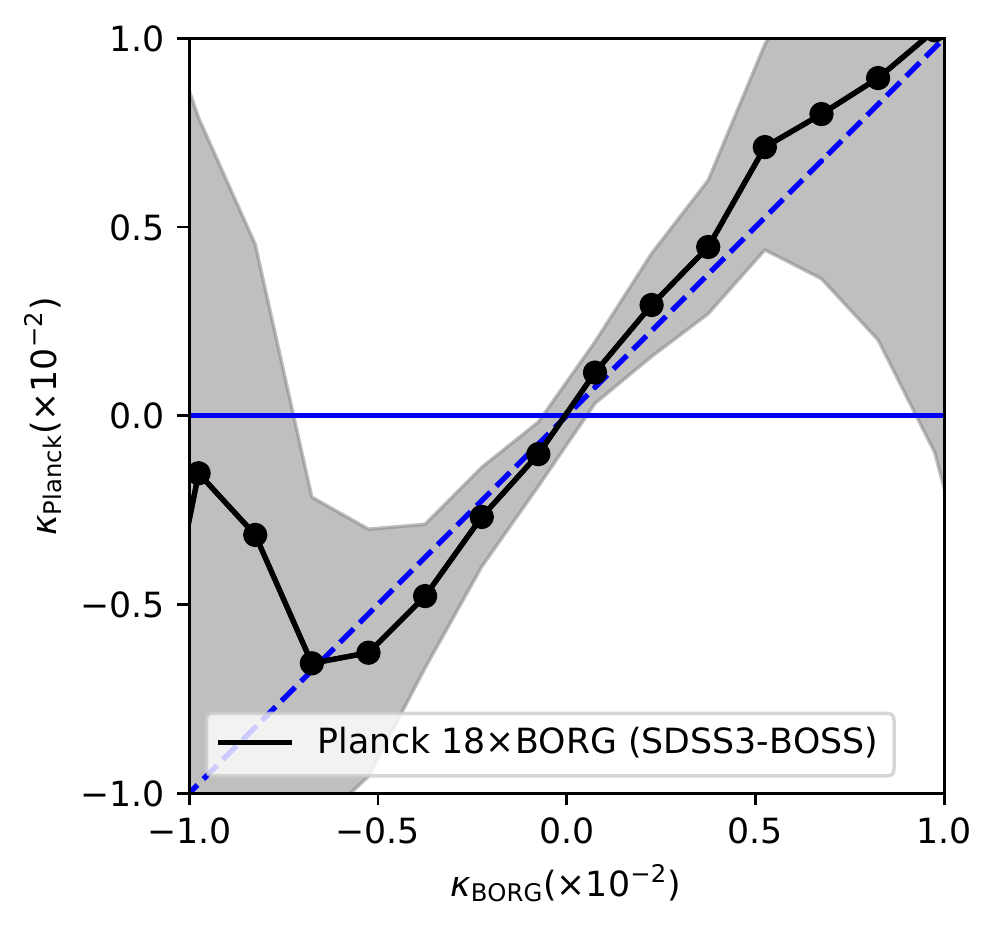}
    \caption{Correlation of the inferred lensing potential from \sdssiii{} and the reconstructed lensing convergence from Planck temperature and polarisation maps. We show here the correlation , for each line-of-sight, between the value of the convergence that is obtained from the Planck lensing map derived in \citet{PlanckCollaboration2018a} and the value that we compute from the gravitational potential derived from \borg{}. Following the procedure used in the Planck analysis, we restricted the modes to $\ell=8-2048$ to derive the pixel based map. The blue dashed line indicates a perfect a correlation. Further details are given in Section~\ref{sec:cross_analysis_cmb}. }
    \label{fig:sdss3_planck_lensing}
\end{figure}

As presented above, the \borg{} algorithm provides very detailed reconstructions of the matter distribution over a cosmological volume. As demonstrated by the analysis of \textit{a posteriori} power spectra in Section~\ref{sec:inferred_density}, inferred initial density fields follow the expected statistical properties at all Fourier modes considered in this work. This is owed to the robust treatment of unknown foreground
effects which could otherwise introduce erroneous features at large scales in the matter distribution. In our previous work \citep[][]{Jasche2019_PM}, we have demonstrated that \borg{} uses the physical forward modelling approach to perform dynamical mass estimates of galaxy clusters. We have found compatible mass profiles with the one derived from gold standard weak lensing measurements, X-ray observations or classical application of the virial theorem to galaxy velocity dispersion. To provide an independent test on whether \borg{} recovered the underlying large-scale dark matter field from \sdssiii{} galaxy clustering data, we here perform a simple cross-correlation analysis with CMB weak lensing data provided by the Planck satellite mission. Achieving this goal requires to first generate posterior templates of weak lensing convergence maps from our inferred mass distributions. We use the classical expression to derive lensing convergence from the matter density fluctuation, which we reproduce here:
\begin{multline}
    \kappa(\hat{n}) = 
    \frac{3}{2} \Omega_\text{m} \left(\frac{H_0}{c}\right)^2 \times \\ \int_{0}^{\chi_\text{CMB}}\frac{\text{d}\chi}{a(\chi)}\; \frac{f_K(\chi) f_K(\chi_\text{CMB}-\chi)}{f_K(\chi_\text{CMB})}\delta_\text{m}(\chi,\hat{n})\;, \label{eq:lensing}
\end{multline}
with $\Omega_\text{m}$ the matter density at redshift $z=0$, $H_0 = 100$~\kmsMpc, $c$ the speed of light, $\chi$ the comoving distance, $f_K(\chi)$ the angular diameter distance, $\chi_\text{CMB}$ the comoving distance to the last scattering surface, $\hat{n}$ the direction in which we observe the convergence. A schematic derivation is provided in Appendix~\ref{app:lensing}.
Having only to rely on the local density fluctuation greatly simplifies the derivation of the convergence map by only taking integrals on lines of sight of the density contrast.

For our cross analysis we use publicly available CMB lensing convergence map\footnote{\url{http://pla.esac.esa.int/pla/}} obtained by the Planck satellite mission \citep[][]{PlanckCollaboration2014a,PlanckCollaboration2018a}.  We have checked that the results are consistent between the 2015 and 2018 maps. The result of the cross-analysis is shown in Figure~\ref{fig:sdss3_planck_lensing}. We show there a direct comparison, for each line of sight, of the convergence computed from the temperature and polarisation maps of the CMB sky observed by the Planck satellite and the one derived from the \borg{} analysis using Equation~\eqref{eq:lensing}. Random samples of the observational noise for the Planck convergence has been taken into account, as well as the fluctuations allowed by the \borg{} posterior constrained by \sdssiii{} data. The correlation procedure automatically cancels out the noise in the lensing map reconstructed from Planck mission data. The grey band is generated by the \borg{} posterior.

Other groups have already reported some correlation between the CMB lensing map obtained by the Planck collaboration and tracers of large-scale scale structures with Sloan Digital Sky Survey data. One of these test is provided in \citet{Singh2017}. However, their comparison between large-scale structures and Planck CMB lensing is done at a much smaller scale than here by focusing on galaxy clusters. Their signal is typically vanishing starting from $\sim$10\Mpch{} from a galaxy, whereas in our case our voxels have a size of $\sim$16\Mpch{}. This shows the future potential of an inferred density map such as the one we are providing, and that we have barely scratched the surface of the amount of available information.
\citet{He2018} attempted a first detection of matter filaments using galaxies of \sdssiii{} through the use of cross-correlation of the angular power spectrum. This detection is however done only at the level of correlation between cross-angular spectra. At larger distances, \citet{Han2019} found some evidence of correlation between the Quasar catalog from SDSS-IV and the same lensing map that we use. Thus we expect that a further extension of the present inference in the SDSS-IV regime would yield even better comparison.

A few other notable examples are the correlation with the CIB-WISE data \citep{Yu2017}, and similarly the correlation with the 2MASS-PhotoZ sample \citep{Bianchini2018}. In these two cases, the sample either covers a larger fraction of the sky or has more galaxies and span a redshift range that is comparable to \sdssiii{}. Additionally \citet{Yu2017} use the CIB contribution which peaks at much farther distances and provide a good template for the lensing convergence, which explains their very high correlation to the Planck lensing map. The WISE component that is used in that work is providing only $\sim$10\% of the correlation. \citet{Bianchini2018} finds also some correlation although it is much weaker owed to the redshift distribution of the galaxies of the 2MPZ which is limited to $z \la 0.2$.

The above agreement is showing that the mass distribution that \borg{} derives from \sdssiii{} is supported by both an independent data-set and an independent physical effect that measure the same quantity. A detailed analysis of CMB-Large scale structure cross analysis will be presented in a forthcoming publication.


\subsection{Mean inferred systematic properties}
\label{sec:inferred_systematic}

\begin{figure*}
    \includegraphics[width=\hsize]{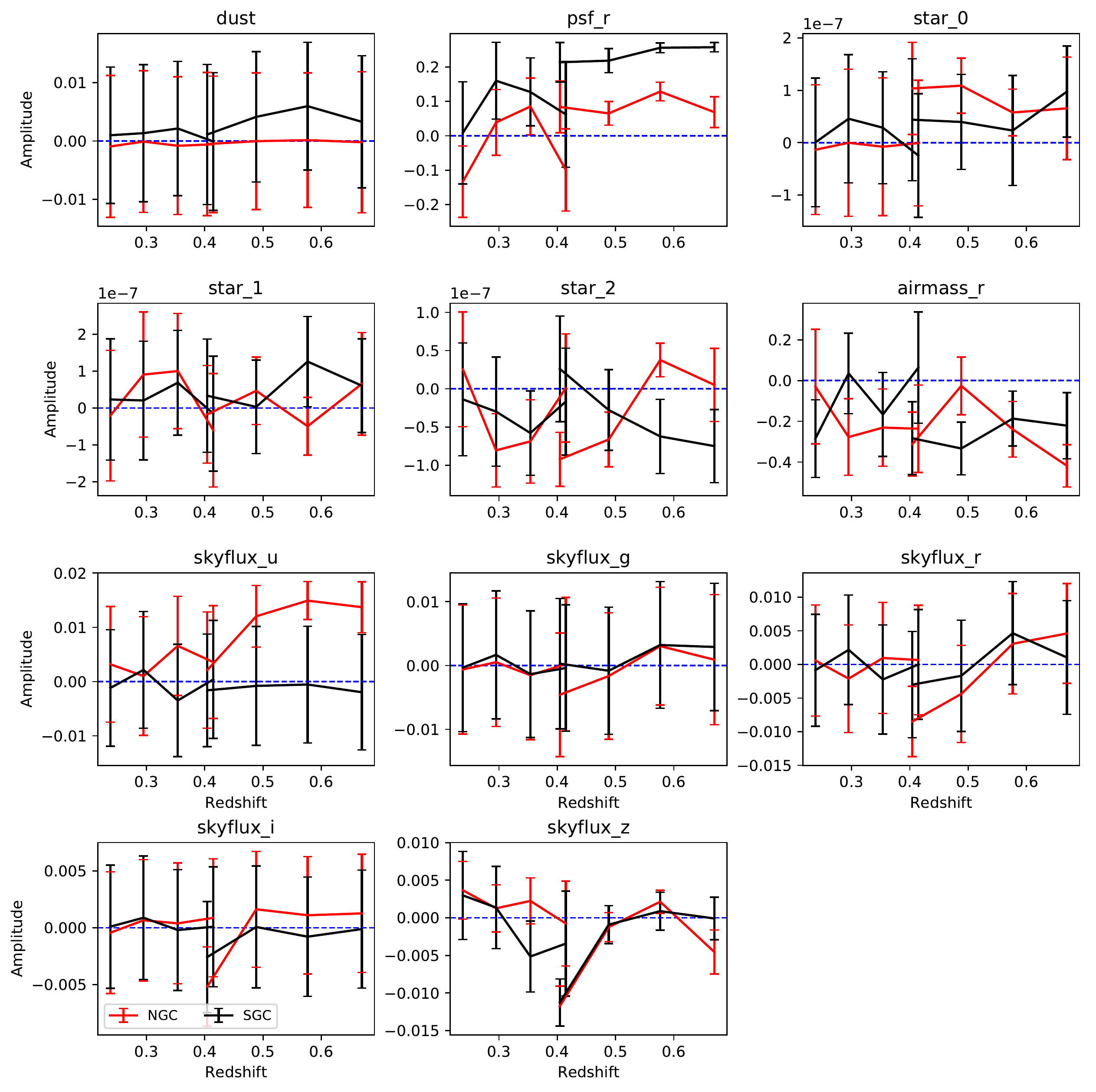}
    \caption{Foreground template coefficients for NGC and SGC as a function of redshift. We show the inferred mean values alongside their standard deviation for each of the 11 foreground templates assigned to each sub-catalogue in our \borg{} run on \sdssiii{}. As for other figures, the LOWZ component is at redshift $z\le 0.4$ and the CMASS component is at $z > 0.4$.}
    \label{fig:foreground_coefficients}
\end{figure*}

In this section, we discuss our results concerning the contamination of the \sdssiii{} sample with known and unknown systematic effects. We remind the reader that we use two techniques at the same time for taking into account these effects, leading to a clean reconstruction of the matter density field: the template based approach \citep[also know as `extended mode' projection, see][]{leistedt_exploiting_2014} and the robust likelihood \citep[closer to `basic mode' projection, see][]{porqueres_explicit_2019}.

\begin{figure*}
    \centering
    \includegraphics[width=\hsize]{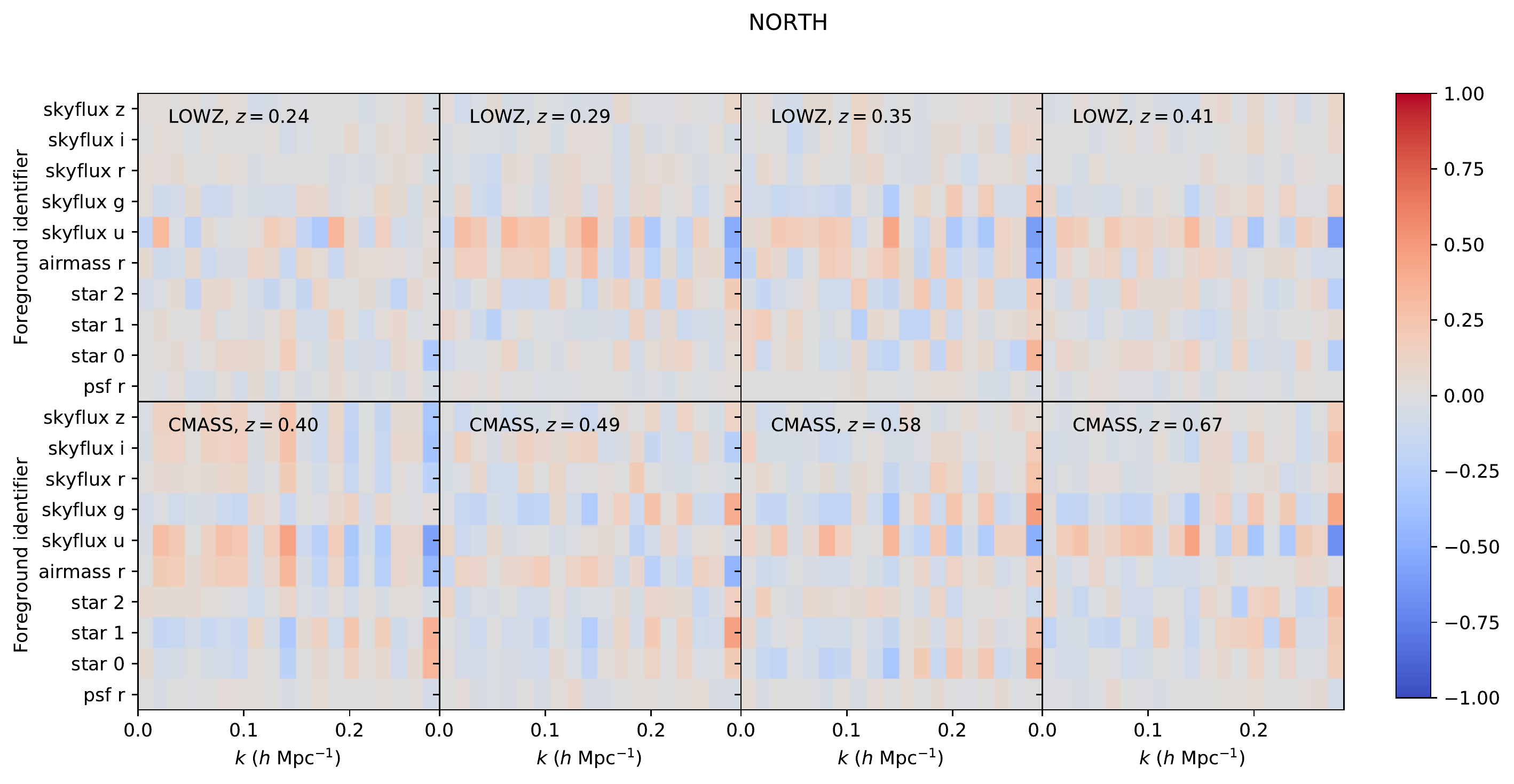}
    \caption{Correlation between foreground template coefficients and the Fourier modes sampled by \borg{} for the NGC sub-samples. The top row contains the correlation matrix for the LOWZ NGC sub-samples ordered from low (\textit{left panel}) to high (\textit{right panel}) redshift. The second row shows the same quantity but for the CMASS NGC sub-samples. }
    \label{fig:foreground_fourier_mode_correlation_ngc}
\end{figure*}

\begin{figure*}
    \centering
    \includegraphics[width=\hsize]{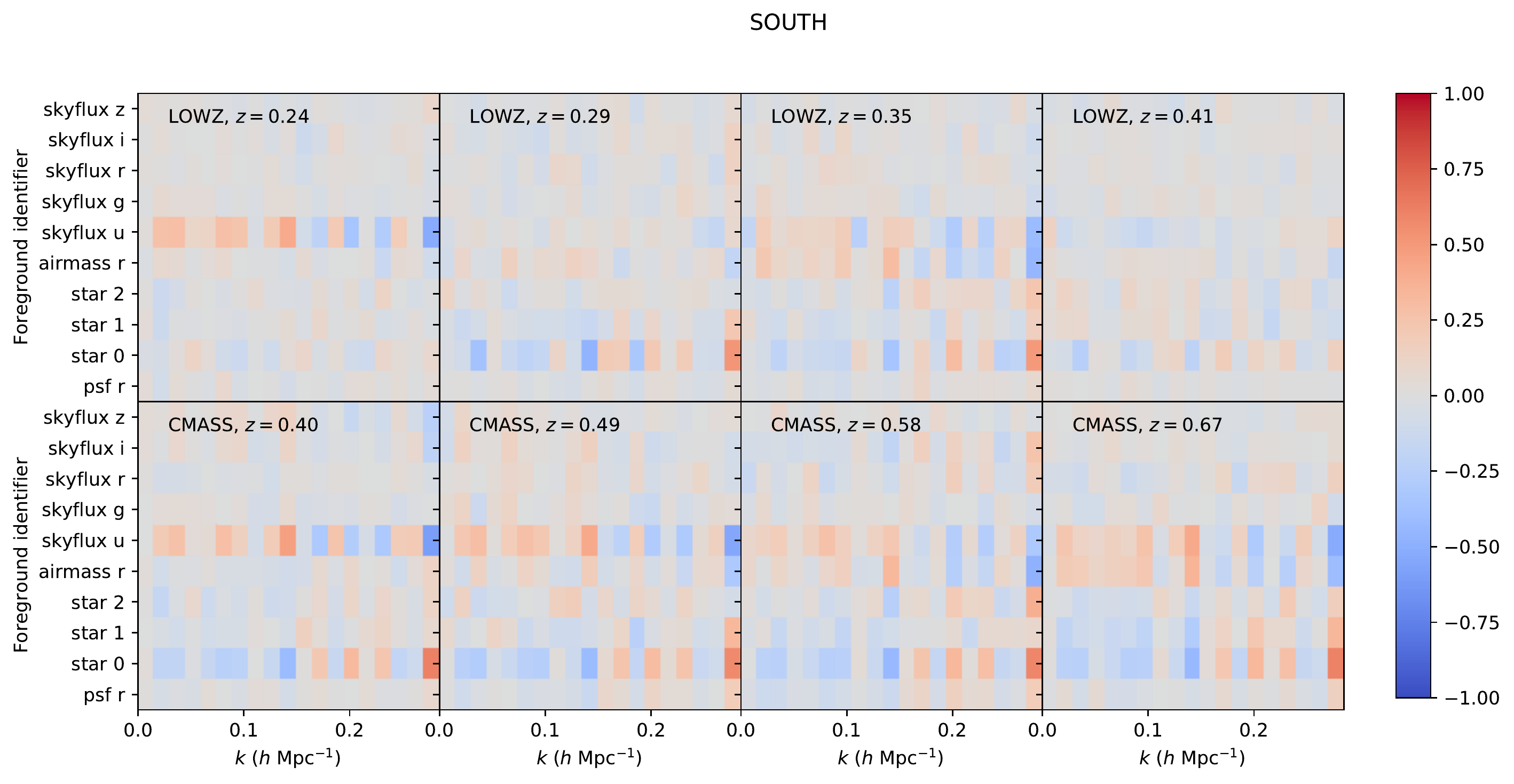}
    \caption{Same as Figure~\ref{fig:foreground_fourier_mode_correlation_ngc} but for the SGC sub-samples.}
    \label{fig:foreground_fourier_mode_correlation_sgc}
\end{figure*}

In Figure~\ref{fig:foreground_coefficients}, we show the mean and standard deviation for each individual foreground coefficients multiplying the indicated templates independently at different redshift and for the NGC and SGC side. 
We note that for a large fraction of these coefficients, no signal is really detectable with our robust likelihood.  For example, the dust contamination is completely flat and compatible with zero. However some of these coefficients exhibit positive, redshift dependent, signal. That is the case for the point spread function (PSF) in the $r$ band (top row, middle panel). In this case there is a clear difference  between NGC and SGC as well. Another template that has clear correlation with data is the skyflux in the $u$ band (third row, left panel). There is a monotonic increase in the contamination level of the \sdssiii{} data in the NGC, while SGC seems to be more immune. We have chosen to use the same choice of foreground templates as the one studied by \citet{ross_clustering_2012}, notably the slices of star density. Though our results are not directly comparable owed to the different procedure to analyse the data, Figure~11 of \citet{ross_clustering_2012} is the most evocative. Generally speaking, this other study showed that the CMASS sample is more contaminated than the LOWZ sample. The star density and the seeing/PSF were among the top contaminant. Here we clearly have evidence of this in the subplot labelled ``psf\_r'' and ``star\_0''. The effect of sky brightness seems larger in our analysis than in the original SDSS analysis. We note that we used the weights provided by the \sdssiii{} collaboration to correct our sample of galaxies before doing the inference, thus some of these contaminations have already been compensated. Our plots may be understood as additional residual contamination that were not accounted for in the galaxy weighting of \sdssiii{}. To conclude this discussion, these results highlight the power of our inference method to detect and correct defects in the data acquisition.

\begin{figure*}
    \centering
    \includegraphics[width=\hsize]{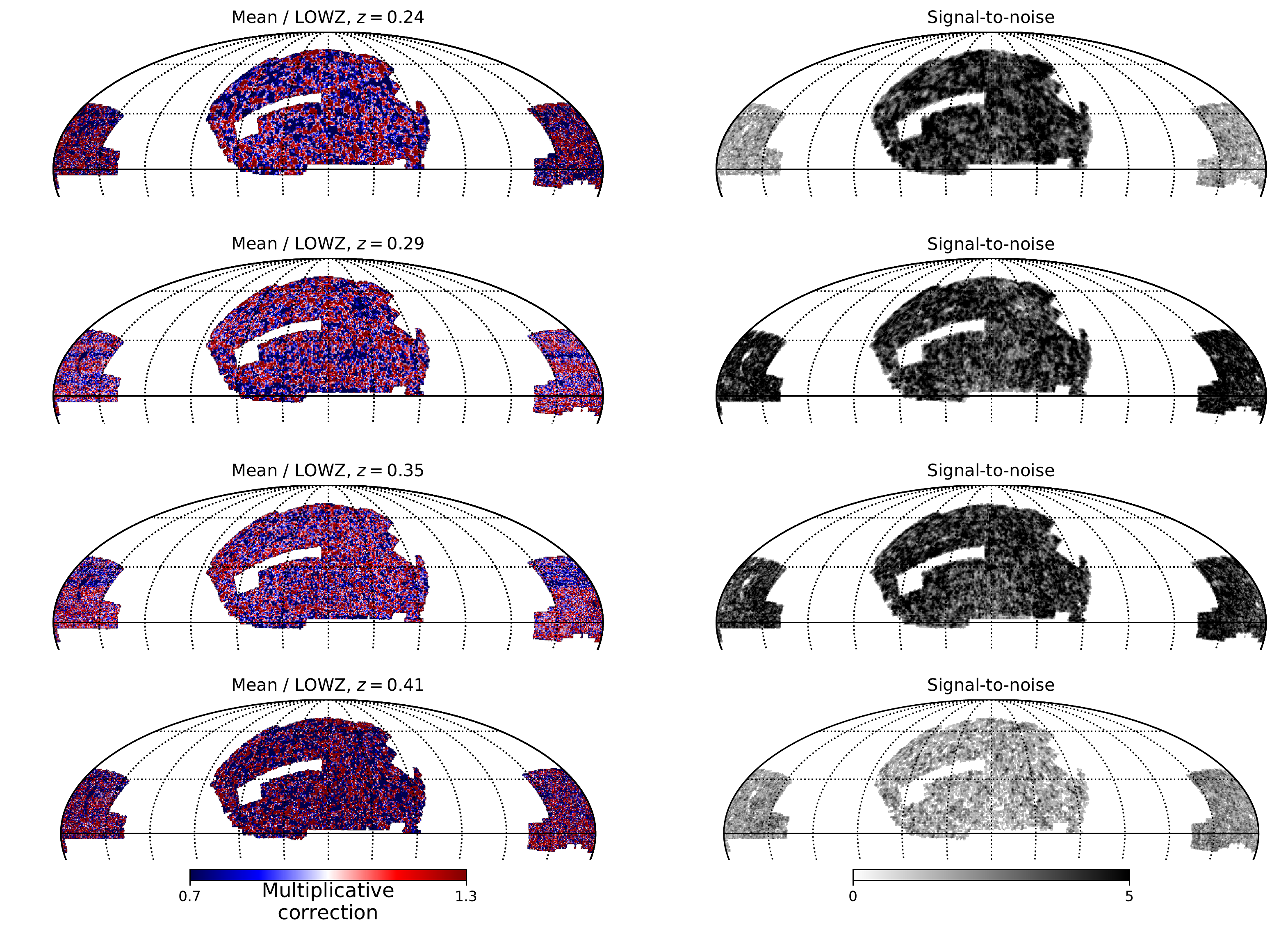}
    \caption{Inferred mean systematic maps (\textit{left column}) as well as the estimated corresponding signal-to-noise for each pixel in these maps (\textit{right panel}), as derived from \sdssiii{} data, in equatorial coordinates. Each of these maps is estimated using Equations~\eqref{eq:mean_robust_map} and  \eqref{eq:variance_robust_map},  independently for each of the redshift bins. The maps are multiplicative, we clearly note a correlated modulation of order 30\% on the sky, indicative of unknown residual systematic effects in the data.  }
    \label{fig:systematic_map_lowz}.
\end{figure*}

\begin{figure*}
    \centering
    \includegraphics[width=\hsize]{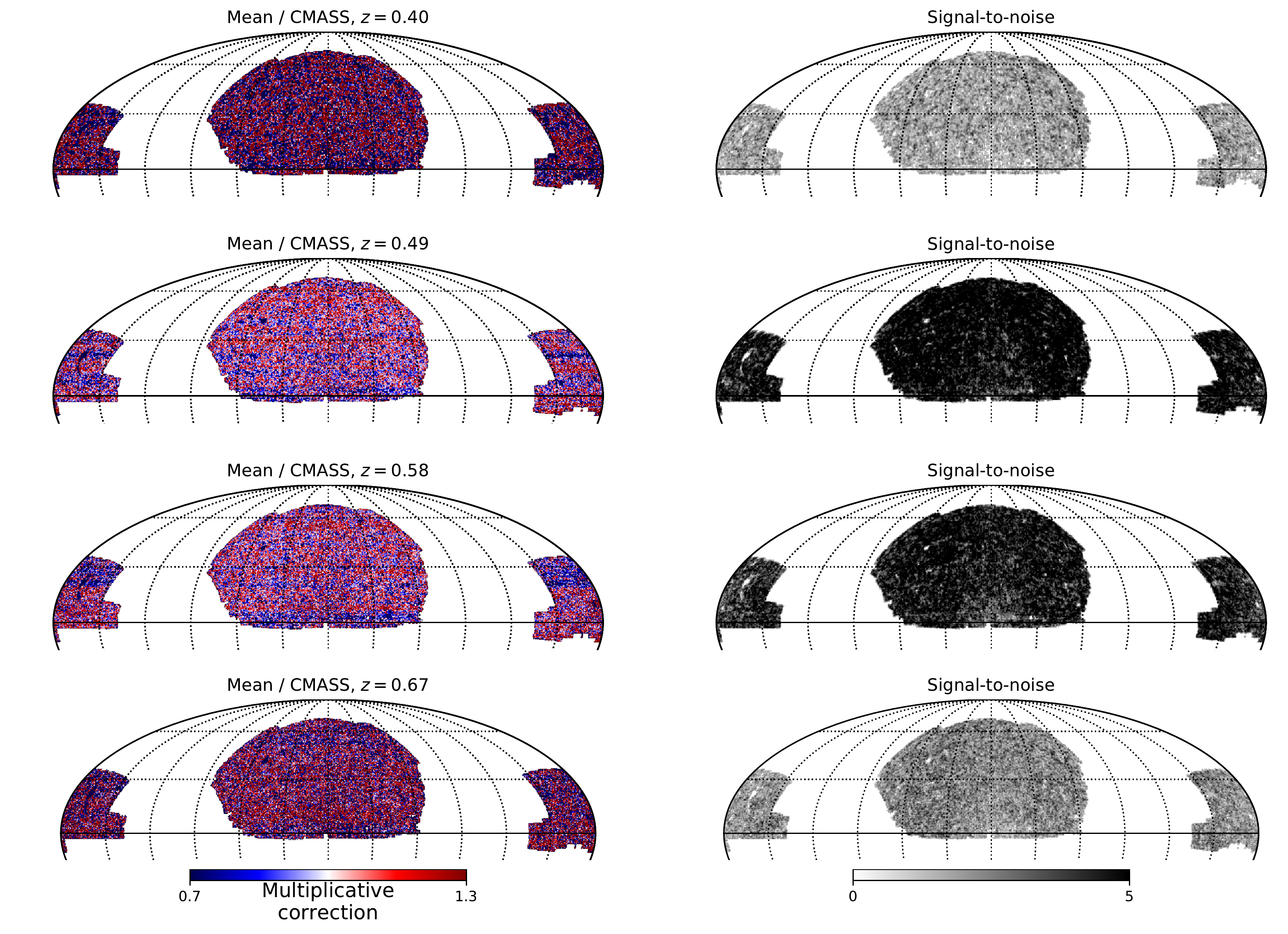}
    \caption{Same as Fig \ref{fig:systematic_map_lowz} but for the CMASS sample.}
    \label{fig:systematic_map_cmass}
\end{figure*}

In Figures~\ref{fig:foreground_fourier_mode_correlation_ngc} and \ref{fig:foreground_fourier_mode_correlation_sgc}, we show the correlation coefficients between these same foreground coefficients and the amplitude of modes at different scales. In both figures, we have ordered in increasing redshift from left-to-right, LOWZ being in the top rows and
CMASS in the bottom rows. The foregrounds templates are indicated on the $y$-axis and the scale on the $x$-axis. We see that despite seeing in most cases a null detection in Figure~\ref{fig:foreground_coefficients}, the amplitudes of coefficients tend to correlate heavily with a lot of modes in the reconstructed initial conditions. This correlation is not stable with redshift nor with scales. The most notable example is the the skyflux in the $u$ band: it is mostly positively correlated with density Fourier modes up to $k\sim 0.15$\hMpc and then becomes negatively correlated at higher $k$, for a lot of sub-catalogues. The influence of the airmass in the $r$ band is also an example of contaminant that changes significantly with redshift in both LOWZ and CMASS. Other foregrounds give an impact that is more focused either spatially or in redshift.

Our Bayesian inference approach has another seducing aspect: unknown foreground contamination may also be reconstructed from posterior samples of the Markov chain, as discussed in Section~\ref{sec:sys_map_inference}. This possibility was already mentioned by \citet{Monaco2018} for Euclid-like surveys. But it is already possible for the \sdssiii{} sample of galaxies. Here we have used equations~\eqref{eq:mean_robust_map} and  \eqref{eq:variance_robust_map} to estimate the ensemble mean and signal-to-noise ratio maps for the four redshift bins of the LOWZ and CMASS samples of \sdssiii{}. The corresponding inference results for unknown foreground contamination are presented in Figures~\ref{fig:systematic_map_lowz} and \ref{fig:systematic_map_cmass}. As can be seen, these maps clearly contain spatial structure despite having marginalised over already 11 foregrounds per sub-catalogue. The systematic maps show clear iso-declination striping, which does not seem to follow the drift-scan strategy of the SDSS photometry. The drift-scan strategy is visible for example in Figure~\ref{fig:systematic_map_prior}. These stripe modulate the signal at the level of 30\% of multiplicative correction on the sky. It is most prominent for the highest signal-to-noise redshift bin at $z=0.29$ and $z=0.35$ for LOWZ and $z=0.49$ and $z=0.58$ for CMASS.

While the striping structure at each redshift bin are fairly represented by some pattern of iso-declination modulation for both NGC and SGC, the pattern itself between the two north and south caps look different. For example in Figure~\ref{fig:systematic_map_lowz} (LOWZ), at $z=0.35$, there is a clear wide blue-stripe ($\sim -30\%$ correction) at DEC$=+30^\circ$ in the SGC, while it is reddish ($\sim +30\%$) for the NGC. It is not strictly iso-declination all the time either. For example in LOWZ, at $z=0.29$ the red stripe at DEC=$+30^\circ$ is widening towards lower DEC while going from left to right of the NGC. Finally, the stripes are not constant with redshift, sometimes inverting completely. That is the case between the two redshift bins of CMASS at $z=0.49$ and $z=0.58$ for which the large stripe just above DEC=$+30^\circ$ is blue in the first case, and red in the second case.

The plausible origin of these systematic effects is likely to be on the ground given the distribution of the stripes. One of such problems are the ``contrails'' \citep{Finkbeiner2016}. Understanding the detail of the origin of these systematic effects is however beyond the scope of this work. The systematic maps will be made available for download on Zenodo after publication.


\section{Conclusion}
\label{sec:conclusion}

With the advent of next-generation galaxy surveys, cosmic large-scale structures will become one of the most important cosmological probes to test the fundamental physics governing the dynamics of our Universe. To ensure continued scientific progress in cosmology, the acquisition of novel quality data needs to be accompanied by the development of novel methods capable of handling unknown systematic effects and to link complex non-linear structure growth physics with observations. 
Such model model of large scale structures as the one we have derived have many  applications for the study and observation of the Universe through different instruments. Some of the applications that were considered in the past are cosmic-web identifications and characterization \citep{LLJW16,Leclercq2017a}, cosmic voids properties \citep{Leclercq2014A}, cosmic magnetic fields \citep{Hutschenreuter2018}, constraints on fifth-force gravity models \citep{desmond_fifth_2018,desmond_fifth_2018-1,desmond_fifth_2019}, peculiar velocity corrections to Hubble-Lemaître constant deduced from standard sirens \citep{Mukherjee2019}.

While traditional methods focus only on analysing a limited number of low-order statistics of the matter distribution, here we apply a fully Bayesian physical forward modelling approach to extract the significant information entailed in the high-order statistics associated to the filamentary matter distribution underlying the galaxies in surveys.

Specifically, we presented a fully Bayesian analysis of the spatial matter distribution probed by \sdssiii{} data. As described in this work, our method infers physically plausible reconstructions from the data while accounting for systematic effects, such as galaxy biases, light-cone effects, survey geometries and other selection effects. Most notably, we demonstrate the application of a novel robust likelihood approach to data, required to deal with unknown systematic effects in the data, which otherwise would result in the erroneous reconstruction of the large-scale matter distribution and corresponding velocity fields, posing significant nuisances for cosmological interpretation of observations.

We conducted an analysis of \sdssiii{} data to recover the cosmic large-scale structure within a Cartesian co-moving volume of 4~000\Mpch{} at a resolution of $\sim$15.6\Mpch{}. Our analysis simultaneously accounts for data in the southern and northern galactic cap of \sdssiii{}. We carefully accounted for non-linear scale dependencies in galaxy biases and data selection effects by splitting the data into galaxy sub-samples of eight redshift bins, nearly equidistant. For each of these galaxy samples, we treated respective systematic effects separately. To model possible non-linear and non-local effects
of the galaxy bias, we proposed a novel multi-power galaxy biasing model, which uses the information of the density field at two different levels of resolution, resulting in a fourteen parameter model per galaxy sub-sample. We determined corresponding bias parameters for each of the galaxy sub-samples, to account for possible redshift evolution. In addition, for each of these galaxy sub-samples, we accounted for survey geometry, and we self-consistently inferred the amplitudes of eleven known foreground templates as well as the unknown noise levels of the galaxy samples. Besides fitting known foreground contributions, a significant improvement over previous work is that our approach uses a robust likelihood approach to also account for unknown systematic effects affecting the survey. As demonstrated in this work, the detailed handling of unknown systematics in galaxy surveys is crucial to infer cosmologically significant and unbiased information from the largest scales in present and coming galaxy surveys. To confirm the reality of the large-scale dynamics that we recovered, we checked the correlation with lensing measurements obtained from the data of the Planck mission. The near-perfect alignment between the prediction that we derived from \sdssiii{} and Planck lensing provides solid evidence that the inferred dark matter density field is correct in the volume spanned by \sdssiii{}.

In summary, the combination of a Bayesian physical forward modelling approach with a robust likelihood approach to account for unknown systematic effects in data is a successful approach to characterise the cosmic large-scale structure and its dynamic formation. The presented work, therefore, defines a promising path towards a fully physically meaningful analysis of next-generation galaxy surveys.

\section*{Acknowledgements}

We thank Fabian Schmidt, Benjamin Wandelt, David Weinberg, François Bouchet, Stéphane Colombi, Valérie de Lapparent, Matthew Lehnert, Suvodip Mukherjee, Peter H. Johansson for useful discussions.
This work has been done within the activities of the Domaine d'Intérêt Majeur (DIM) ``Astrophysique et Conditions d'Apparition de la Vie'' (ACAV), and received financial support from Région Ile-de-France. GL acknowledges financial support from the ILP LABEX, under reference ANR-10-LABX-63, which is financed by French state funds managed by the ANR within the programme ``Investissements d'Avenir'' under reference ANR-11-IDEX-0004-02. GL also acknowledges financial support from the ANR BIG4, under reference ANR-16-CE23-0002. FL acknowledges funding from the Imperial College London Research Fellowship Scheme. 
This  work  was granted  access  to  the  HPC  resources  of  CINES  (Centre  Informatique National de l'Enseignement Sup\'erieur) under the allocation A0020410153 and A0040410153  made by GENCI and  has  made  use  of  the  Horizon  cluster  hosted  by  the  Institut  d'Astrophysique  de  Paris  on  which  the  cosmological simulations  were post-processed. GL thanks the hospitality of the University of Helsinki where part of this work took place. This work is done within the Aquila Consortium\footnote{\url{https://www.aquila-consortium.org/}}.




\bibliographystyle{mnras}
\bibliography{sdss} 

\appendix

\section{Variance of the displacement field}
\label{app:displacement}

In a $\Lambda$CDM universe, assuming that evolution of large-scale structures is well described by the Zel'Dovich approximation \citep{ZelDovich1970}, the statistics of the displacement is simple. Using the continuity equation, we can write
\begin{equation}
    \nabla_{\mvec{q}} . \mvec{\Psi} = -D(t) \delta({\mvec{q}}),
\end{equation}
with $\mvec{q}$ the Lagrangian coordinates, which at high redshift are close to the Eulerian coordinates, $D(t)$ the growth function, $\mvec{\Psi}$ the displacement field. The one-point variance of the displacement field becomes thus
\begin{align}
    \langle \Psi_a^2(t,\mvec{q}) \rangle &= 
    \int \frac{\text{d}^3 \mvec{k} \text{d}^3 \mvec{k}'}{(2\pi)^6} \mathrm{e}^{i (\mvec{k}+\mvec{k}').\mvec{q}} \langle \hat{\Psi}_a(\mvec{k})\hat{\Psi}_a(\mvec{k}')\rangle  \nonumber \\
    &= D^2(t) \int \frac{\text{d}^3 \mvec{k} \text{d}^3 \mvec{k}'}{(2\pi)^6} \frac{-k_a k'_a}{|\mvec{k}|^2 |\mvec{k}'|^2} \mathrm{e}^{i (\mvec{k}+\mvec{k}').\mvec{q}} \langle \hat{\delta}(\mvec{k})\hat{\delta}(\mvec{k}')\rangle \nonumber \\
    &= D^2(t) \int  \frac{\text{d}^3\mvec{k}}{(2\pi)^3} P(k) \frac{k_a^2}{k^4} \nonumber \\
    &= \frac{D^2(t)}{3}\int \frac{\text{d}^3\mvec{k}}{(2\pi)^3} \frac{1}{k^2} P(k) \\
    &= \frac{D^2(t)}{12\pi^2}\int_{k=0}^{+\infty} \text{d}k P(k)\;,
\end{align}
with $P(k)$ the power spectrum of matter density fluctuations at high redshift. For a $\Lambda$CDM universe, with Planck 2018 cosmology, the square root of that variance is 5.96\Mpch{}. An acceptable typical upper bound to the displacement field may be at $\sim$3 times that value, which leads to 17.9\Mpch{}.

\section{Adjoint gradient of the bias model}
\label{app:ag_bias}

Computing of the adjoint gradient, or back-propagation in machine learning terminology, consists in linearly transforming an error vector back to the adequate parameter space of interest. In \borg, that consists in transporting the error vector from the likelihood space, which touches galaxy distribution, to the initial condition. The bias model step relate the matter density to the expected galaxy distribution, before the effect of the pipeline of detection by the instrument. We assume that we are provided an error vector $v_i$, per mesh element. The new error vector $\tilde{v}_q$ will be derived as follow:
\begin{equation}
    \tilde{v}_q = \sum_{i} v_i \frac{\partial N^{(g)}_i}{\partial \delta_q} = 2 \sum_i v_i \frac{\partial \mvec{\Delta}_i}{\partial \delta_q}^\dagger \mmat{Q} \mvec{\Delta}_i.
\end{equation}
In general the element of the vector $\mvec{\Delta}_i$ take the following form
\begin{equation}
    \left(\mvec{\Delta}_i\right)_a = (\delta^{(\ell_a)}_i)^{\gamma_a},
\end{equation}
with $j_a$ the density averaging level at the component $a$ and $\gamma_a$ the power rising of the component $a$. The detail of that ordering is given in Section~\ref{sec:galaxy_bias}. The special case $j_a = 0$ corresponding to $\delta^{(0)} = 1$.
Thus we may derive the derivative of the vector $\mvec{\Delta}_i$ by looking at each component:
\begin{align}
    \frac{\partial \Delta_{i,a}}{\partial \delta_q} & = \frac{\partial \delta^{(j_a)}_i}{\partial \delta_q} \times
        \left\{\begin{array}{ll}
            \gamma_a \left(\delta^{(j_a)}_i\right)^{\gamma_a-1}, & \text{if } \gamma_a \ge 1, j_a \ge 1 \\
            0 & \text{otherwise}
          \end{array}\right. \\
          & = \frac{\partial \delta^{(j_a)}_i}{\partial \delta_q} g_a\left(\delta^{(j_a)}_i\right)
\end{align}
Finally the derivative of the averaging operator is
\begin{equation}
    \frac{\partial \delta^{(\ell_a)}_i}{\partial \delta_q} =
    \frac{1}{8^\ell} \times \left\{ \begin{array}{ll}
        1 & \text{if } q \in \mathcal{V}^{(j_a)}_i\;, \\
        0 & \text{otherwise,}
    \end{array}\right.
\end{equation}
with $\mathcal{V}^{(j_a)}_i$ the vicinity set of $i$ at the level $j_a$ of the oct-tree. This vicinity set is defined implicitly from Equation~\eqref{eq:coarsening}. We may compute it explicitly by doing the matrix-vector multiplication with the vector $\mvec{v}$:
\begin{multline}
  \tilde{v}_q = \frac{1}{8^\ell} 
  \sum_{\alpha,\beta} \sum_{a,b,c=0}^{2^{\ell-1}} v_{f_\ell(q,a,b,c)} \times \\ g_\alpha\left(\delta^{(\ell_\alpha)}_{f_\ell(q,a,b,c)}\right) Q_{\alpha,\beta} \left(\delta^{(\ell_\alpha)}_{f_{\ell(q,a,b,c)}}\right)^{\gamma_\alpha}.
\end{multline}
This gives an explicit algorithm to compute the adjoint-gradient with this new bias model.

\section{Lensing equation}
\label{app:lensing}

In this appendix we give a brief reminder of the derivation of Equation~\eqref{eq:lensing}. If we consider the Newtonian potential $\Psi$ defined at comoving distances $\chi$ and angular direction $\hat{\bf{n}}$ on the sky \citep{Kaiser1998,Lewis2006,Kilbinger2015}, then the sky displacement of one photon, at first order of perturbation in $\Psi$ and on the geodesic trajectory followed by that photon is:
\begin{multline}
    \mvec{\alpha}(\mvec{\hat{n}}) =  \frac{1}{c^2} \times \\
     \int_0^{\chi_{\mathrm{CMB}}} \mathrm{d}\chi\;
    \frac{f_K(\chi_{\mathrm{CMB}}-\chi)}{f_K(\chi_\text{CMB}) f_K(\chi)} (\mvec{\nabla}_{\hat{n}} \Psi)(\chi \mvec{\hat{n}}; \chi_{\mathrm{CMB}}-\chi)
\end{multline}
where $\chi$ is the comoving radial distance and
\begin{equation}
    f_K(\chi)= \left\{\begin{array}{ll}
        \sin(\chi)  & \text{for }K=+1\text{, closed universe;} \\
        \chi        & \text{for }K=0\text{, flat universe;} \\
        \sinh(\chi) & \text{for }K=-1\text{, open universe}.
    \end{array}\right.
\end{equation}
The convergence is defined as the sky divergence of the sky displacement:
\begin{equation}
    \kappa(\mvec{\hat{n}}) =  \mvec{\nabla}_{\hat {n}} \mvec{\alpha}(\mvec{\hat{n}}),
\end{equation}
Furthermore, the three-dimensional potential $\Psi$ is related to the matter density contrast $\delta_m(\mvec{x})$ via the Poisson equation in comoving coordinates,
\begin{equation}
    \nabla_{\mvec{x}}^2 \Psi = \frac{3}{2 a(\chi)} \Omega_m(\chi) H^2_0  \delta_{\mathrm{m}}(\mvec{x};\chi)\;.
\end{equation}
The above equation is valid in the usual perturbative regime of the metric, which is the case for the entirety of this work.
By moving the divergence inside the integral, we obtain
\begin{multline}
    \kappa(\mvec{\hat{n}}) = -\frac{1}{c^2} \times \\
    \int_0^{\chi_{\mathrm{CMB}}}     \mathrm{d}\chi\;
    \frac{f_K(\chi_{\mathrm{CMB}}-\chi) f_K(\chi)}{f_K(\chi_{\mathrm{CMB}})} \times \\ \left((\mvec{\nabla}^2_{\hat{n}} \Psi)(f_K(\chi) \mvec{\hat{n}}; \chi_\text{CMB} - \chi)\right) \;,
\end{multline}
As generally done in the scientific literature and explicitly justified in \citet{Kilbinger2015}, we replace the 2D Laplacian by the 3D Laplacian because we expect the second-order radial derivatives to average to zero at the scale that we consider. 
Thus we have a simplified expression for the convergence
\begin{multline}
    \kappa(\hat{n}) = \frac{3}{2} \Omega_m \left(\frac{H_0}{c}\right)^2 \times \\ \int_{0}^{\chi_{CMB}}\text{d}\chi\; \frac{f_K(\chi) f_K(\chi_{CMB}-\chi)}{f_K(\chi_{CMB})}\frac{\delta_m(\chi,\hat{n})}{a(\chi)}
\end{multline}
This greatly simplifies the derivation of the convergence map by only 
taking integrals on line of sights of the density contrast.

\section{Testing the warm-up phase of the sampler}
\label{app:warmup}

\begin{figure}
    {
        \centering
        \includegraphics[width=\hsize]{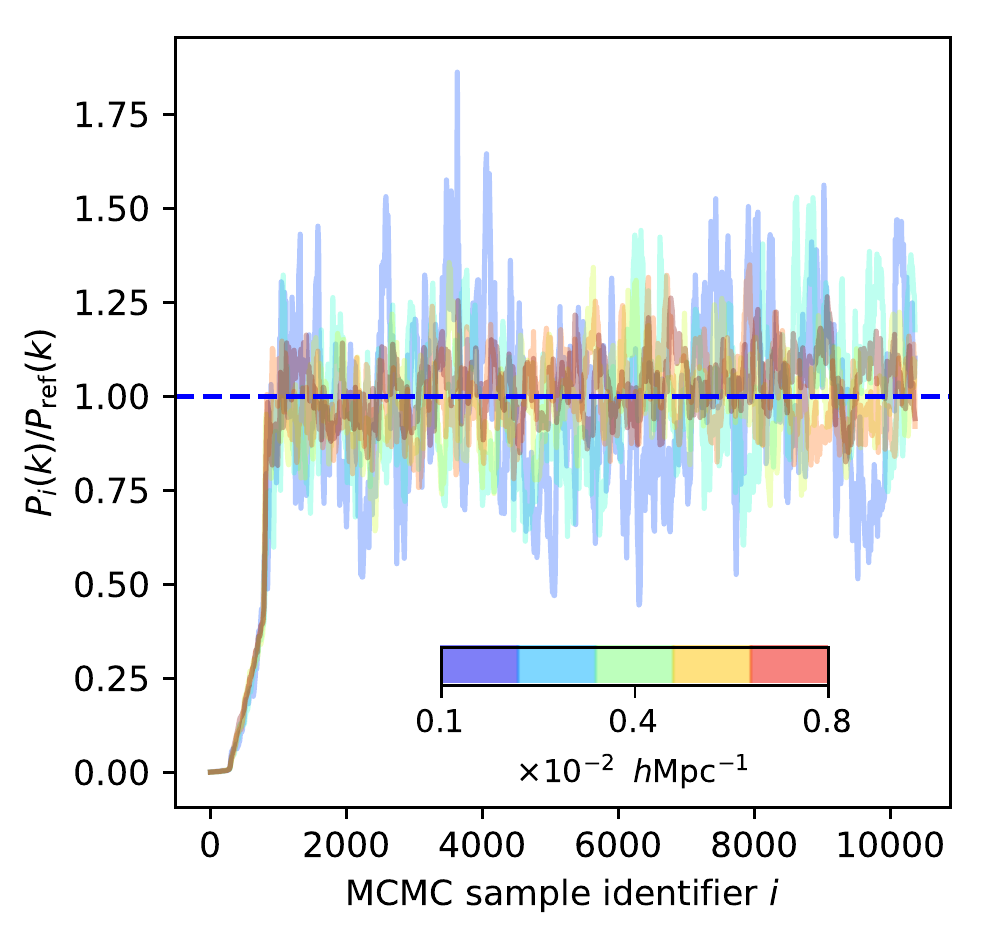}
    }
    \caption{Amplitudes of the \textit{a posteriori} primordial matter power spectrum at different Fourier modes traced during the warm-up phase of the MCMC sampler. As can be seen, initially, modes perform a coherent drift towards the high probability region in posterior distribution and start oscillating around their fiducial values once the Markov chain has reached a stationary state.}
    \label{fig:burnin_Pk}
\end{figure}

As described in our previous works \citep{jasche_bayesian_2013,lavaux_unmasking_2016,Jasche2019_PM}, we initialize the Markov chain with an over-dispersed state, that is far remote from the target regions in the parameter space. This permits us to test the sampler behaviour during the initial warm-up phase and confirm it has approached the stationary regime before starting to record Markov samples for the analysis.
Over-dispersed initial states are prepared by initialising the Markov chain with a random Gaussian initial density field scaled by a factor $1/10$, which translates to $1/100$ in Figure~\ref{fig:burnin_Pk}. To follow the sampler behaviour during its warm-up phase, we follow the traces of posterior power spectrum amplitudes throughout the initial sampler steps. As demonstrated by Figure~\ref{fig:burnin_Pk}, initially power spectrum amplitudes at the different modes of Fourier-space perform a coherent drift towards preferred regions in parameter space. After about $1,000$ Markov transition steps the chain has reached a stationary distribution and power spectrum amplitudes oscillate around their expected fiducial values. From that moment, we start recording samples from the stationary distribution to perform the analysis presented in this work.

\bsp	
\label{lastpage}
\end{document}